\input harvmac.tex 
\input amssym.def
\input amssym
\input epsf.tex
\magnification\magstep1
\baselineskip 14pt

\parskip 6pt
\newdimen\itemindent \itemindent=32pt
\def\textindent#1{\parindent=\itemindent\let\par=\resetpar%
\indent\llap{#1\enspace}\ignorespaces}

\let\oldpar=\par
\def\resetpar{\oldpar\parindent=20pt\let\par=\oldpar}

\font\ninerm=cmr9 \font\ninesy=cmsy9
\font\eightrm=cmr8 \font\sixrm=cmr6
\font\eighti=cmmi8 \font\sixi=cmmi6
\font\eightsy=cmsy8 \font\sixsy=cmsy6
\font\eightbf=cmbx8 \font\sixbf=cmbx6
\font\eightit=cmti8
\def\eightpoint{\def\rm{\fam0\eightrm}
  \textfont0=\eightrm \scriptfont0=\sixrm \scriptscriptfont0=\fiverm
  \textfont1=\eighti  \scriptfont1=\sixi  \scriptscriptfont1=\fivei
  \textfont2=\eightsy \scriptfont2=\sixsy \scriptscriptfont2=\fivesy
  \textfont3=\tenex   \scriptfont3=\tenex \scriptscriptfont3=\tenex
  \textfont\itfam=\eightit  \def\it{\fam\itfam\eightit}%
  \textfont\bffam=\eightbf  \scriptfont\bffam=\sixbf
  \scriptscriptfont\bffam=\fivebf  \def\bf{\fam\bffam\eightbf}%
  \normalbaselineskip=9pt
  \setbox\strutbox=\hbox{\vrule height7pt depth2pt width0pt}%
  \let\big=\eightbig  \normalbaselines\rm}
\catcode`@=11 %
\def\eightbig#1{{\hbox{$\textfont0=\ninerm\textfont2=\ninesy
  \left#1\vbox to6.5pt{}\right.\n@@space$}}}
\def\vfootnote#1{\insert\footins\bgroup\eightpoint
  \interlinepenalty=\interfootnotelinepenalty
  \splittopskip=\ht\strutbox %
  \splitmaxdepth=\dp\strutbox %
  \leftskip=0pt \rightskip=0pt \spaceskip=0pt \xspaceskip=0pt
  \textindent{#1}\footstrut\futurelet\next\fo@t}
\catcode`@=12 %
\def \de{\delta}

\def \tr{{\rm tr }}

\def \hf{{f}}

\def \T{{\rm T}}

\def \hG{{\hat {\cal G}}}
\def \hH{{\hat {\cal H}}}

\def \l{\big \langle}
\def \r{\big \rangle}

\def \half{{\textstyle {1 \over 2}}}

\def \quar{{\textstyle {1 \over 4}}}
\def \ts{\textstyle}

\def \d{{\rm d}}

\def \A{{\cal A}}
\def \B{{\cal B}}
\def \C{{\cal C}}

\def \F{{\cal F}}
\def \G{{\cal G}}
\def \H{{\cal H}}
\def \I{{\cal I}}

\def \K{{\cal K}}

\def \M{{\cal M}}
\def \N{{\cal N}}

\def \Q{{\cal Q}}
\def \S{{\cal S}}

\def \vphi{{\varphi}}

\def \oD{{\overline D}}

\def \bx{\bar x}
\def \hb{\hat b}
\font \bigbf=cmbx10 scaled \magstep1


\lref\Witt{E. Witten, {\it Anti-de Sitter space and holography},
Adv. Theor. Math. Phys. 2 (1998) 253, hep-th/9802150.}
\lref\Free{D.Z. Freedman, S.D. Mathur, A. Matsusis and L. Rastelli, 
{\it Correlation functions in the CFT${}_d$/AdS${}_{d+1}$ correspondence}, Nucl. 
Phys. B546 (1999) 96, hep-th/9804058\semi
D.Z. Freedman, S.D. Mathur, A. Matsusis and L. Rastelli, {\it Comments on 4-point 
functions in the CFT/AdS correspondence}, Phys. Lett. B452 (1999) 61, 
hep-th/9808006\semi
E. D'Hoker and D.Z. Freedman, {\it General Scalar Exchange in AdS${}_{d+1}$}, 
Nucl. Phys. B550 (1999) 261, hep-th/9811257\semi
E. D'Hoker and D.Z. Freedman, {\it General Boson Exchange in $AdS_{d+1}$},
Nucl. Phys. B544 (1999) 612, hep-th/9809179\semi
E. D'Hoker, D.Z. Freedman, S.D. Mathur, A. Matsusis and L. Rastelli, 
{\it Gravity and gauge boson propagators in $AdS_{d+1}$}, 
Nucl. Phys. B562 (1999) 330, hep-th/9902042\semi
E. D'Hoker, D.Z. Freedman and L. Rastelli, {\it AdS/CFT 4-point functions: How to
succeed at z-integrals without really trying}, Nucl. Phys. B562 (1999) 395, 
hep-th/9905049.}

\lref\FreeD{E. D'Hoker, D.Z. Freedman, S.D. Mathur, A. Matsusis and 
L. Rastelli, {\it Graviton exchange and complete four-point functions in the
AdS/CFT  correspondence}, Nucl. Phys. B562 (1999) 353, hep-th/9903196.}
\lref\Hok{E. D'Hoker, S.D. Mathur, A. Matsusis and L. Rastelli, {\it The operator 
product expansion of N = 4 SYM and the four-point functions of 
supergravity}, Nucl. Phys. B589 (2000) 38, hep-th/9911222.}

\lref\Sei{S. Lee, S. Minwalla, M. Rangamani and N. Seiberg, {\it Three-Point
Functions of Chiral Operators in $D=4$, $\N=4$ SYM at Large $N$},
Adv. Theor. Math. Phys.  2 (1998) 697, hep-th/9806074.}

\lref\AFa{G.E.~Arutyunov and S.A.~Frolov, {\it Quadratic action for type IIB 
supergravity on $AdS_5 \times S^5$}, JHEP 9908 (1999) 024, hep-th/9811106.}
\lref\AFb{
G.~Arutyunov and S.~Frolov, {\it Some cubic couplings in type IIB 
supergravity on $AdS_5 \times S^5$ and three-point functions in SYM${}_4$ at 
large N}, Phys.\ Rev.\ D61 (2000) 064009, hep-th/9907085.}
\lref\AFc{
G.~Arutyunov and S.~Frolov, {\it Scalar quartic couplings in type IIB 
supergravity on $AdS_5 \times S^5$}, Nucl.\ Phys.\ B579 (2000) 117, 
hep-th/9912210.}
\lref\Lee{
S.M.~Lee, {\it AdS${}_5$/CFT${}_4$ four-point functions of chiral primary 
operators: Cubic vertices}, Nucl.\ Phys.\ B563 (1999) 349, 
hep-th/9907108.}
\lref\AFd{
G.~Arutyunov and S.~Frolov, {\it Four-point functions of lowest weight CPOs 
in N = 4 SYM(4) in supergravity approximation}, Phys.\ Rev.\ D62 
(2000) 064016; hep-th/0002170.}
\lref\ADHS{G. Arutyunov, F.A. Dolan, H. Osborn and E. Sokatchev,
{\it Correlation Functions and Massive Kaluza-Klein Modes in the AdS/CFT
Correspondence}, Nucl. Phys. B665 (2003) 273, hep-th/0212116.}
\lref\Degen{G. Arutyunov and E. Sokatchev, {\it On a Large N Degeneracy
in $\N=4$ SYM and the AdS/CFT Correspondence}, Nucl. Phys. B663 (2003) 163,
hep-th/0301058.}

\lref\DO{F.A. Dolan and H. Osborn, {\it Conformal Four Point Functions and the
Operator Product Expansion}, Nucl. Phys. B599 (2001) 459,  hep-th/0011040.}
\lref\DH{F.A. Dolan and H. Osborn, {\it Conformal Partial Waves and the
Operator Product Expansion}, Nucl. Phys. B678 (2004) 491,  hep-th/0309180.}

\lref\Intr{K. Intriligator, {\it Bonus Symmetries of ${\cal N} =4$ 
Super-Yang-Mills Correlation Functions via AdS Duality}, 
Nucl. Phys. B551 (1999) 575, hep-th/9811047.}
\lref\Bon{K. Intriligator and W. Skiba, {\it Bonus Symmetry and the Operator
Product Expansion of ${\cal N} =4$ Super-Yang-Mills},
Nucl. Phys. B559 (1999) 165, hep-th/9905020.}

\lref\NO{M. Nirschl and H. Osborn, {\it Superconformal Ward Identities and
their Solution}, Nucl. Phys. B711 (2005) 409, hep-th/0407060.}
\lref\SCFT{F.A. Dolan and H. Osborn, {\it Superconformal Symmetry, Correlation 
Functions and the Operator Product Expansion}, Nucl. Phys. B629 (2002) 3,
hep-th/0112251.}
\lref\OPEexp{F.A. Dolan and H. Osborn, {\it Conformal Partial Wave Expansions for
$\N=4$ Chiral Four Point Functions}, Ann. Phys. to be published, hep-th/0412335.}
\lref\OPEN{G. Arutyunov, S. Frolov and A.C. Petkou, {\it Operator Product
Expansion of the Lowest Weight CPOs in $\N=4$ SYM${}_4$ at Strong Coupling},
Nucl. Phys. B586 (2000) 547, hep-th/0005182;
(E) Nucl. Phys. B609 (2001) 539.}
\lref\OPEW{G. Arutyunov, S. Frolov and A.C. Petkou, {\it Perturbative and
instanton corrections to the OPE of CPOs in $\N=4$ SYM${}_4$}, Nucl.
Phys. B602 (2001) 238, hep-th/0010137; (E) Nucl. Phys. B609 (2001) 540.}

\lref\Short{F.A. Dolan and H. Osborn, {\it On short and semi-short
representations for four-dimensional superconformal symmetry}, Ann. Phys.
307 (2003) 41, hep-th/0209056.}
\lref\bpsN{B. Eden and E. Sokatchev, {\it On the OPE of $1/2$ BPS Short
Operators in $N=4$ SCFT${}_4$}, Nucl. Phys. B618 (2001) 259, hep-th/0106249.}
\lref\Hes{P.J. Heslop and P.S. Howe, {\it OPEs and 3-point correlators of
protected operators in $N=4$ SYM}, hep-th/0107212.}
\lref\Howe{P.J. Heslop and P.S. Howe, {\it Four-point functions in $N=4$ SYM},
JHEP 0301 (2003) 043, hep-th/0211252.}

\lref\Edent{B. Eden, A.C. Petkou, C. Schubert and E. Sokatchev, {\it Partial
non-renormalisation of the stress-tensor four-point function in $N=4$
SYM and AdS/CFT}, Nucl. Phys. B607 (2001) 191, hep-th/0009106.}
\lref\Except{G. Arutyunov, B. Eden, A.C. Petkou and E. Sokatchev,
{\it Exceptional non-renorm-alization properties and OPE analysis of
chiral four-point functions in $\N=4$ SYM${}_4$},
Nucl. Phys. B620 (2002) 380, hep-th/0103230.}
\lref\ASok{F.A. Dolan, L. Gallot and E. Sokatchev, {\it On Four-point Functions
of $\half$-BPS operators in General Dimensions}, JHEP 0409 (2004) 056,
hep-th/0405180.}
\lref\Gub{S.S. Gubser, I.R. Klebanov and A.M. Polyakov, {\it Gauge Theory 
Correlators from Non-Critical String Theory}, Phys. Lett. B428 (1998) 105,
hep-th/9802109.}

\lref\index{J. Kinney, J. Maldacena, S Minwalla and S Raju, {\it An Index
for 4 dimensional Super Conformal Theories}, hep-th/0510251.}

{\nopagenumbers
\rightline{DAMTP/06-8}
\rightline{hep-th/0601148}
\vskip 1.5truecm
\centerline {\bigbf Conjectures for Large $N$ $\N=4$ Superconformal}
\vskip 4pt
\centerline {\bigbf Chiral Primary Four Point Functions}
\vskip  6pt
\vskip 2.0 true cm
\centerline {F.A. Dolan, M. Nirschl and H. Osborn${}^\dagger$}
\vskip 12pt
\centerline {\ Department of Applied Mathematics and Theoretical Physics,}
\centerline {Wilberforce Road, Cambridge CB3 0WA, England}
\vskip 1.5 true cm

{\eightpoint
\parindent 1.5cm{

{\narrower\smallskip\parindent 0pt

An expression for the four point function for $\half$-BPS operators
belonging to the $[0,p,0]$ $SU4)$ representation in $\N=4$ superconformal 
theories at strong coupling in the large $N$ limit is suggested for any $p$. 
It is expressed in terms of the four point integrals defined by integration
over $AdS_5$ and agrees with, and was motivated by, results for $p=2,3,4$
obtained via the AdS/CFT correspondence. Using crossing symmetry and unitarity,
the detailed form is dictated by the requirement that at large $N$ the 
contribution of long multiplets with twist less than $2p$, which do not have 
anomalous dimensions, should cancel corresponding free field contributions.

PACS no: 11.25.Hf; 11.30.Pb

Keywords: Conformal field theory, Operator product expansion, Four point
function, Large N, Superconformal Symmetry.

\narrower}}

\vfill
\line{${}^\dagger$ 
address for correspondence: Trinity College, Cambridge, CB2 1TQ, England\hfill}
\line{\hskip0.2cm emails:
{\tt fad20@damtp.cam.ac.uk}, {\tt michael.nirschl@gmail.com}\hfill}
\line{\hskip 1.3cm  and {\tt ho@damtp.cam.ac.uk}\hfill}
}

\eject}
\pageno=1

\newsec{Introduction}

The discovery of the AdS/CFT correspondence has demonstrated the intimate
relationship between string theory and quantum field theory. In particular
the initial proposal of an essential equivalence of type IIB string theory
on $AdS_5 \times S^5$ with the maximal $\N=4$ superconformal gauge theories
on the four dimensional boundary has proved especially fruitful. The isometry
group of $AdS_5 \times S^5$,  $SO(4,2) \times SO(6) \simeq SU(2,2) \times SU(4)$,
matches precisely the bosonic part of the $\N=4$ superconformal group $PSU(2,2|4)$.
For gauge group $SU(N)$ and a gauge coupling $g$ then at large $N$ the 
supergravity approximation to ten dimensional string theory may be used to 
determine the strong coupling, $\lambda = g^2 N \to \infty$, behaviour of
the $\N=4$ superconformal gauge theory. As shown by Witten \Witt\ results for the
leading large $N$ behaviour of correlation functions of gauge invariant operators
can be calculated by considering Feynman graphs involving propagators on $AdS_5$
linking points on the $AdS_5$ boundary to supergravity vertices in the bulk.

For $\N=4$ superconformal theories the simplest operators to discuss are
the so called chiral primary operators which are annihilated by half the
supercharges and so are part of short $\half$-BPS multiplets.
These are scalars belonging to the $SU(4)_R$ representation with Dynkin
labels $[0,p,0]$ and have a protected scale dimension $\Delta=p$. They
are simply related to Kaluza-Klein modes in the expansion of supergravity
fields on $S^5$. The three point functions of such operators were determined
by Lee {\it et al} \Sei, see also \AFb. In this case the $x$-dependence is 
determined by conformal invariance apart from an overall constant. The 
normalisation constant calculated for large $N$ using supergravity in the 
strong coupling limit is identical to the result for free field theory for 
large $N$, which corresponds to the expectation that the three point functions 
for chiral primary operators are independent of $g$ \refs{\Intr,\Bon}. The 
corresponding four point functions however have a non trivial dependence on 
the coupling, although this is also constrained by non renormalisation theorems 
\refs{\Edent,\Except} and superconformal Ward identities \refs{\Howe,\NO,\ASok}. 
This is reflected by the fact that the operator product expansion allows for the
presence of contributions from long multiplets which have anomalous dimensions
expressible perturbatively as an expansion in $g$. 

Using the AdS/CFT correspondence 
the basic building blocks for $n$-point correlation functions are,
as defined in \FreeD, given by integrals on $AdS_{d+1}$ of the form
\eqn\defD{
D_{\Delta_1 \dots \Delta_n}(x_1,\dots,x_n) = 
{1\over \pi^{{1\over 2}d}} \int_0^\infty \!\!\! \d z \int \d^d x \,
{1\over z^{d+1}} \, \prod_{i=1}^n
\bigg ({z \over z^{2} + (x-x_i)^2}  \bigg )^{\! \Delta_i }\, ,
}
where $x_i$ are points on the boundary $S^d$ of $AdS_{d+1}$. The $n$-point 
functions defined by \defD\ transform covariantly under conformal
transformations on the $x_i$ with corresponding scale dimension $\Delta_i$. 
Manifestly they are symmetric under permutations of $x_i, \Delta_i$. For $n=3$ the
integral can be evaluated to give the standard conformal form for the three point
function for operators of scale dimension $\Delta_i$.  For $n=4$ it may reduced to
a function $\oD_{\Delta_1\Delta_2\Delta_3\Delta_4}(u,v)$, independent of $d$, of
two conformal invariants $u,v$. In addition to relations inherited from
the permutation symmetry there are various non trivial identities relating 
$\oD$-functions with $\Delta_i$ differing by integers and $\sum \Delta_i$
varying by $\pm 2$, \refs{\FreeD,\DO}.
Calculations of amplitudes on $AdS_5$ with four points on the boundary, involving
scalar, vector and graviton exchanges, may be reduced to linear combinations
$D_{\Delta_1\Delta_2\Delta_3\Delta_4}(x_1,x_2,x_3,x_4)$ for various integer 
$\Delta_i$, and also $\sum_i \Delta_i$ even \Free.

These methods then allow the determination of the large $N$ strong coupling limit
for  chiral primary operator four point functions. The simplest case, and the 
first which was determined, is the four point function for $[0,2,0]$ $\half$-BPS 
operators \AFd. These highly non trivial calculations,
which require expanding supergravity to fourth order in fluctuations around $AdS_5
\times S^5$ \refs{\AFa,\AFc,\Lee}, were later extended to obtain the four point 
functions for chiral 
primary operators belonging  to the $[0,p,0]$ representation for $p=3,4$ 
\refs{\ADHS,\Degen}.
The supergravity results generate contributions from disconnected diagrams
which are ${\rm O}(1)$ in the large $N$ limit and which correspond to the
disconnected pieces in free field theory and also, from connected diagrams,
${\rm O}(1/N^2)$ pieces which reproduce the results of free field theory to
this order. There is in addition a dynamical contribution which is a linear 
combination of $\oD$-functions. Using $\oD$-identities is necessary to show 
that these results are in accord with the consequences of superconformal 
symmetry. For $p=2$, superconformal symmetry requires that the dynamical part 
may be reduced to a single crossing symmetric function of $u,v$ and, although 
not initially evident this  can be simplified to a single $\oD$ function,
$\oD_{2422}(u,v)$, which is essentially crossing symmetric as a consequence
of $\oD_{2422}(u,v)=\oD_{2422}(v,u)=\oD_{2422}(u/v,1/v)/v^4$.  For $p=3,4$
the dynamical part reduces respectively to 1,2 functions of $u,v$ together with 
their transformations related by crossing (the group of crossing transformations
on 4-point functions $\S_3$ is here generated by $u\leftrightarrow v$ and 
$u\to u/v, \, v \to 1/v$).  The number of $\oD$-functions in the final 
answer proliferate. Nevertheless the results for $p=3,4$ can be reduced to 
expressions in which  they are all just of the form $\oD_{i\, p+2\, jk}$ for 
various $i,j,k \le p$ \NO.

The $\oD$ functions have a representation as a series expansion in powers
of $u,1-v$ but in which terms proportional to $\ln u$ are also present. The 
part involving $\ln u$ is interpreted as arising from the leading term in 
the $1/N$ expansion of the anomalous dimensions of long multiplets whose 
contributions involve factors $u^\Delta$. Anomalous dimensions are possible
only for long multiplets which are here denoted by $\A^\Delta_{nm,\ell}$,
where the lowest dimension operator belongs to a $SU(4)$ representation having
Dynkin labels $[n-m,2m,n-m]$ with also scale dimension $\Delta$, spin $\ell$.
In the operator product expansion of the four point function for $[0,p,0]$ 
chiral primary operators long multiplets may only be present for $m\le n \le p-2$ 
\refs{\bpsN,\Hes}. This also follows from superconformal Ward identities \NO.
The conformal partial wave analysis of the strong coupling results for
four point functions demonstrates
that anomalous dimensions are obtained \refs{\OPEN,\SCFT,\ADHS,\OPEexp} only
for long multiplets $\A^\Delta_{nm,\ell}$ which have twist $\Delta-\ell \ge 2p$.
However the unitarity bound in superconformal representation theory requires just 
$\Delta - \ell \ge 2n +2$. Assuming all long multiplets necessarily have non
zero anomalous dimensions to leading order in $1/N$ these results then require 
that long multiplets with twist $\Delta-\ell < 2p$ are absent in the operator 
product expansion for two $[0,p,0]$ chiral primary operators in the large $N$ 
limit. 

Contributions without anomalous dimensions correspond to operators belonging to 
short $\B_{nm}$ or semi-short superconformal multiplets $\C_{nm,\ell}$ whose scale 
dimensions are protected, the lowest dimension operator belonging to the
$SU(4)$ representation $[n-m,2m,n-m]$ has $\Delta = n$, with $\ell=0$, or 
$\Delta=n+2+\ell$ respectively. The semi-short multiplets are
related to the decomposition of long multiplets at the unitarity threshold
\Short
\eqn\longm{
\A^{2n+\ell+2}_{nm,\ell}  \simeq \C_{nm,\ell} \oplus \C_{n+1\, m,\ell-1} 
\oplus \dots \, ,   \qquad 0\le m \le n \, ,
}
neglecting two additional semi-short multiplets which do not contribute to the 
operator product expansion of two $[0,p,0]$ chiral primary operators. The
relation \longm\ extends to $\ell=0$ if we identify \Short
\eqn\shortm{
\C_{nm,-1} \simeq \B_{n+1\, m}  \, ,
}
where for $n>m$ $\B_{nm}$ is a $\quar$-BPS multiplet. Only such short
or semi-short multiplets therefore contribute to the operator product
expansion of the four point function considered here in the large
$N$ limit for twist $\Delta-\ell<2p$. 

As a consequence of \longm\ there is a potential ambiguity in the
decomposition of the operator product expansion into contributions from
various supermultiplets in a theory in which long multiplets do not
have anomalous dimensions. Following \index\ if we denote $n[\M]$ the
number of supermultiplets $\M$ contributing to the operator product
expansion for two $[0,p,0]$ chiral primary operators then the index
\eqn\ind{
I_{nm} = \sum_{\ell=-1}^{n-m} (-1)^{\ell+1} \, n[ \C_{n-\ell\, m , \ell} ] \, ,
}
is such that combinations of short or semi-short multiplets forming a long
multiplet cancel. The number of $\half$-BPS multiplets $n[\B_{nn}]$ and
also $n[\B_{n+1\, n}]$ are also invariants. In the case of interest here
by crossing multiplets $\M_{nm,\ell}$ can only contribute if $n+m+\ell$
is even (in consequence $\quar$-BPS multiplets $\B_{n+1\, n}$ cannot appear)
so that $I_{nm}$ is relevant only for $n+m$ even. The results obtained
from analysis of the four point function for the operator product 
expansion for two $[0,p,0]$ chiral primary operators to order $1/N$ is that 
$I_{nm}\ne 0$ for all $0\le m \le n$ with $n\ge p-2$ and $I_{nm}=0$ for $n<p-2$. 
The index is saturated by requiring just the multiplets $\C_{p-1\, m,\ell}$
and $\B_{pm}$ to be present. There are as well $\half$-BPS multiplets
$\B_{nn}$ for $n=1,\dots p$.

{}From the large $N$ results for the $p=2$ case it was shown in \OPEN\ that the
only twist two singlet operator necessary in the operator product expansion
is when $\ell=2$, corresponding to the energy momentum tensor. This was
confirmed in \SCFT\ where the absence of all leading twist two singlet operators
belonging to long supermultiplets was also demonstrated for any $\ell$
using a simplified form of the large $N$ results.
This is as expected since these operators are absent in the large $N$ limit, 
their dimensions are proportional to $\lambda^{1\over 4}$ as 
$\lambda \to \infty$ \Gub\ so that they decouple from the spectrum revealed 
by the large $N$ operator product expansion.
This is different from the perturbative expansion about free field theory
when such operators are present for any $\ell$ ($\ell=0$ is the Konishi scalar)
and have a leading anomalous dimension proportional to $\lambda$, which is
therefore not suppressed by $1/N$, whereas multi-trace operators have anomalous 
dimensions which are ${\rm O}(\lambda/N^2)$ \OPEW. The disappearance of twist 
two operators 
belonging to long multiplets in the strong coupling limit $\lambda \to \infty$ 
requires in the operator product calculations in \SCFT\ a non trivial cancellation
between the free field ${\rm O}(1/N^2)$ contributions and also the leading
non $\ln u$ terms from the dynamical $\oD$-functions. These cancellations
were shown to extend also to the $p=3,4$ cases in \OPEexp. The operator
product expansion for large $N$ strong coupling then has contributions only
from multi-trace operators with  anomalous dimension suppressed by $1/N^2$.

We use this cancellation in this paper as a guiding principle to determine an 
expression
for the four point function of single trace $[0,p,0]$ chiral primary operators 
for any $p$. The free field results are straightforward to obtain to order
$1/N^2$. Although there are many, ${[{1\over 12}p(p+6)]+1}$, crossing symmetric 
forms, each of which may in general have a different multiplicative constant 
depending on $N$ in a complicated fashion, as $N\to \infty$ then 
these simplify to just $p^2\over N^2$ or $2p^2\over N^2$, assuming
the coefficients of disconnected contributions are normalised to 1.
The factor $p^2$ reflects the cyclic symmetry of the trace for
single trace operators. The dynamical part at large $N$ is assumed to be
expressible as a linear combination of $\oD$-functions.
In  order to accommodate the consequences of $SU(4)_R$ symmetry we introduce, 
instead of rank $p$ symmetric traceless tensors for each $[0,p,0]$ chiral
primary field, variables $\sigma,\tau$ so that the four
point amplitude becomes a polynomial in $\sigma,\tau$ of degree $p$. The number
of independent terms corresponds exactly to the number of invariants formed
by four $[0,p,0]$ representations, or the number of irreducible 
representations in the tensor product $[0,p,0] \otimes [0,p,0] \simeq
\bigoplus_{m=0,\dots ,n, \, n = 0, \dots ,p} [n-m,2m,n-m]$. The operator
product expansion for four point function is then equivalent to a
simultaneous expansion in terms of harmonic functions of $\sigma,\tau$ for
$SO(6)$ and also in harmonic functions of $u,v$ for the non compact group
$SO(4,2)$. Crossing
symmetry transformations are now generated by $u \leftrightarrow v , 
\, \sigma \to \sigma/\tau, \tau \to 1/\tau$,  and $u\to u/v, \, v \to 1/v, \,
\sigma \leftrightarrow \tau$. With the aid of superconformal symmetry the 
dynamical part reduces to a crossing symmetric form $\H$ which is a polynomial
in $\sigma,\tau$ of degree $p-2$ \NO. The operator product expansion for $\H$
corresponds only to long multiplets. We determine all possible crossing
symmetric expressions for this dynamical term formed from 
$\oD_{i\, p+2\, jk}(u,v)$ for  appropriate $i,j,k$, which are constrained by 
removal of identities amongst such
$\oD$-functions and also by the requirement that in each channel unitarity
constraints on operators appearing in the operator product expansion are
satisfied. Although each crossing symmetric form has $\ln u$ terms which
arise in the operator product expansion only for $\Delta-\ell \ge p$
there are also sub leading terms which contribute for $\Delta - \ell <p$.
Our assumption is that this must be cancelled by a corresponding free
contribution. Within our framework this determines all coefficients
uniquely, independent of any supergravity calculations on the basis of
the AdS/CFT correspondence. Of course this agrees with known results for
$p=2,3,4$. 

In detail in the next section we define the four point correlation functions
which are investigated here and introduce six dimensional $SO(6)$ null
vectors in terms of which the variables $\sigma,\tau$ are constructed. We
also give the two variable harmonic functions $Y_{nm}(\sigma,\tau)$ which
allow a decomposition into $SU(4)$ representations. In section three we
recapitulate expressions for the two variable conformal partial waves, 
which are harmonic functions $u,v$ with a similar form to $Y_{nm}$, and
show how the expansion can be simplified by first considering a decomposition
into contributions for given twist which is described by a single variable
function of a single variable $x$. Section four uses this and superconformal
identities to show a decomposition which matches the contributions of
differing superconformal multiplets in the operator product expansion. The
results are equivalent to those in \NO, as used in \OPEexp, but are in
a form more convenient for later discussion here. Section five gives the results
for free field theory, initially in terms of crossing symmetric polynomials
but then in terms of  the simplified form valid for large $N$. It is shown
how contributions involving twist less than $2p$ are universal in that
they do not depend on the particular $p$, apart from a common overall
coefficient. Expressions for the universal
functions of $x$ are obtained which are rather complicated but which are
simplified in some cases. In section six we analyse the possible
crossing symmetric forms for $\oD$-functions with appropriate additional
constraints as described above. The general form for the dynamical part
of the four point function is then supposed to be an arbitrary linear
combination of the independent allowed expressions. We also exhibit
those contributions from the $\oD$-functions which play a crucial role in
our discussion. These are expressed as a linear combination of hypergeometric
functions of $1-v$ and are required to have a universal form. The constraints
arising from this are considered in section seven for low $p$ and it is
shown how this matches known results. A general analysis which gives a unique
solution is presented in section eight. The derivation of the results in
sections 6,7,8 does not depend substantially on the rest of this paper and may 
be read independently. In section 9 we demonstrate the cancellation between 
the universal parts of the dynamical and free contributions to the four point
function which is the fundamental principle behind the proposals in this
paper. A few remarks are made in a final conclusion.

Some details are deferred to three appendices. In appendix A we amplify
some of the results of section 4 giving detailed contributions of short
and semi-short $\N=4$ superconformal multiplets to the conformal
partial wave expansion. In appendix B we  calculate the large $N$ limit
for the free field contributions to the four point function which is used
in section 5. In appendix C we prove some necessary hypergeometric identities.

\newsec{Superconformal Correlation Functions}

We here establish the essential notation for the discussion of four point
correlation functions for chiral primary $\half$-BPS operators, spinless with 
scale dimension $\Delta=p$ and belonging to the
$SU(4)$ $R$-symmetry $[0,p,0]$  representation, which are the subject of
interest in this paper.  These  chiral primary operators belong to the simplest 
short multiplets of $\N=4$ superconformal symmetry and are represented  by 
symmetric traceless $SO(6)$ tensor fields 
$\vphi_{r_1 \dots r_p}(x)$, with $r_i =1,\dots,6$. For detailed analysis for 
arbitrary $p$ it is very convenient to consider instead 
$\vphi^{(p)}(x,t) = \vphi_{r_1\dots r_p}(x)\, t_{r_1} \dots t_{r_p}$, homogeneous 
of degree $p$ in $t$,  for $t_r$ an arbitrary six dimensional complex null vector.
Such null vectors were employed in \ADHS\ and used in detail for simplifying
superconformal transformations in \NO.
The four point correlation functions of chiral primary operators may then be 
written in the form
\eqn\fourp{
\l \vphi^{(p)} (x_1,t_1) \, \vphi^{(p)}(x_2,t_2)\,
\vphi^{(p)}(x_3,t_3)\, \vphi^{(p)}(x_4,t_4 )\r
=  \bigg ( {t_1 {\cdot t_2} \, t_3 {\cdot t_4} \over
x_{12}^{\, 2}\, x_{34}^{\, 2} } \bigg )^{\! p} \G^{(p)}(u,v;\sigma,\tau) \, ,
}
with the definitions
\eqn\defuv{
x_{ij} = x_i - x_j \, , \qquad
u = { x_{12}^{\, 2}\, x_{34}^{\, 2} \over x_{13}^{\, 2}\, x_{24}^{\, 2} }
\, ,\quad
v = { x_{14}^{\, 2}\, x_{23}^{\, 2} \over x_{13}^{\, 2}\, x_{24}^{\, 2} }
\, ,
}
and $\sigma,\tau$ $SU(4)$ invariants which are homogeneous of degree zero and
are defined by
\eqn\defst{
\sigma = {t_1{\cdot t_3} \, t_2{\cdot t_4} \over
t_1{\cdot t_2} \, t_3{\cdot t_4}} \, , \qquad
\tau  = {t_1{\cdot t_4} \, t_2{\cdot t_3} \over
t_1{\cdot t_2} \, t_3{\cdot t_4}} \, .
}
Necessarily, since the correlation function is homogeneous of degree $p$
in each $t_i$,  $\G^{(p)}(u,v;\sigma,\tau)$ is a polynomial of degree $p$ 
in $\sigma, \,\tau$ (i.e it may be expanded in the $\half(p+1)(p+2)$ 
monomials $\sigma^r \tau^s$ with $r+s\le p$). Crossing symmetry requires
\eqn\crossG{
\G^{(p)}(u,v;\sigma,\tau) = \G^{(p)}(u/v,1/v;\tau,\sigma) =
\Big ( {u\over v} \Big )^p \tau^{p}\, \G^{(p)} (v,u;\sigma/\tau,1/\tau) \, .
}
As related in the introduction above, for $p=2,3,4$ expressions for each term in the
$SU(4)$ expansion of $\G^{(p)}(u,v;\sigma,\tau)$ have been obtained for large
$N$  through the AdS/CFT correspondence.

For the subsequent discussion it is necessary to consider new
variables $x,\bx$ defined by
\eqn\defxx{
u = x \bx \, , \qquad v = (1-x)(1-\bx) \, .
}
In an analogous fashion to \defxx\ we may also write
\eqn\defaa{
\sigma = \alpha \bar \alpha  \, , \qquad \tau = (1-\alpha )(1- \bar \alpha) \, ,
}
and $\G^{(p)}(u,v;\sigma,\tau)$ may be written instead in terms of
a symmetric function of $x,\bx$ and also $ \alpha, \bar \alpha$. We may
decompose this into contributions for the different $SU(4)$ representations
formed by the tensor product $[0,p,0] \otimes [0,p,0]$ by writing
\eqn\Gnm{
\G^{(p)}(u,v;\sigma,\tau) = \sum_{0\le m \le n \le p} \G^{(p)}_{nm}(u,v)
\, Y_{nm}(\sigma,\tau) \, ,
}
where $Y_{nm}$ are two variable harmonic polynomials of degree $n$ which
correspond to the $SU(4)$ representation with Dynkin labels ${[n-m,2m,n-m]}$.
Explicitly we have
\eqn\poly{
Y_{nm}(\sigma,\tau) =
{P_{n+1}(y) P_{m}(\bar y) - P_{m} (y) P_{n+1} (\bar y)
\over y - \bar y} \, , \quad  y = 2\alpha-1 \, , \ \bar y = 2 \bar \alpha -1\, ,
}
with $P_n$ the usual Legendre polynomials. Using the standard results
\eqn\Leg{
\half yP_n(y) = \gamma_{n,1} P_{n+1}(y) + \gamma_{n,-1} P_{n-1}(y) \, , \qquad
\gamma_{n,1} = {n+1 \over 2(2n+1)} \, , \quad 
\gamma_{n,-1} = {n\over 2(2n+1)} \, ,
}
we easily obtain the following recurrence
relations which play a significant role in the later discussion,
\eqn\recurY{\eqalign{
(\sigma - \tau) Y_{nm}(\sigma,\tau) = {}& \! \sum_{r=\pm 1} \big (
\gamma_{n+1,r} Y_{n+r \, m}(\sigma,\tau) + 
\gamma_{m,r} Y_{n\, m+r }(\sigma,\tau)\big ) \, , \cr
\half (\sigma + \tau - \half ) Y_{nm}(\sigma,\tau) = {}& \! \sum_{r,s=\pm 1} 
\gamma_{n+1,r} \gamma_{m,s} \, Y_{n+r \, m+s }(\sigma,\tau) \, .  \cr}
}
Here we note that if appropriate we should take $Y_{n\, n+1}=0$ and
$Y_{n-1\, n+1} = - Y_{nn}$.

\newsec{Conformal Partial Waves}

Using the operator product expansion the four point function for scalar 
conformal primary fields may be expanded in terms of contributions from 
conformal primary operators with scale dimension $\Delta$ and spin $\ell$ 
giving a conformal partial wave expansion. This is equivalent to an
expansion in terms of functions $G^{(\ell)}_\Delta(u,v)$ of the two conformal 
invariants $u,v$ which are essentially harmonic functions for $SO(4,2)$ \DH.
In a perturbative context we are then interested in determining 
the coefficients $a_{j,\ell}$ for an expansion of the form \DO
\eqn\cexp{
F(u,v) = \sum_{j,\ell=0,1,\dots} \!\! 
a_{j,\ell} \, u^j G^{(\ell)}_{2a+2j+\ell} (u,v) \, ,
}
where we require that $F(u,v)$ has a power series expansion in $u,1-v$. 
In  four dimensions the conformal partial waves are
expressed simply in terms of the variables $x,\bx$  given by \defxx\ by
\eqn\defG{
u^j G^{(\ell)}_{2a+2j+\ell} (u,v) = - {g_{a,j+\ell+1}(x)\, g_{a,j}(\bx)
-  g_{a,j}(x)\, g_{a,j+\ell+1}(\bx)  \over x - \bx } \, ,
}
where
\eqn\defg{
g_{a,j}(x) = (-x)^j F(a+j-1,a+j-1;2a+2j-2;x) \, , \quad j=0,1,2,\dots \, .
}
which satisfies $g_{a,j+n}(x) = (-x)^n g_{a+n,j}(x)$.
As a consequence of the form \defG\ we may impose
\eqn\rela{
a_{j,\ell} = - a_{j+\ell+1,-\ell-2} \, , \qquad a_{j,-1} = 0 \, , 
}
and then the expansion \cexp\ may be written in the form
\eqn\cxexp{
(x-\bx) F(u,v) = \sum_{j=0}^\infty F_j(x) \, g_{a,j}(\bx) \, ,
}
where
\eqn\Fexp{
F_j(x) = \sum_{\ell=0}^{j-1} \! a_{j-\ell-1,\ell} \, g_{a,j-\ell-1}(x)
- \sum_{\ell=0}^\infty \! a_{j,\ell} \, g_{a,j+\ell+1}(x) 
= - \sum_{\ell=-j-1}^\infty \! a_{j,\ell} \, g_{a,j+\ell+1}(x) \, .
}

As special cases we may note that 
\eqn\spec{\eqalign{
G^{(\ell)}_{\ell} (u,v) = {}& - {(1-\half \bx) \, g_{0,\ell+1}(x)
-  (1-\half x)\, g_{0,\ell+1}(\bx)  \over x - \bx } \, , \cr
u G^{(\ell)}_{\ell+2} (u,v) = {}&  {\bx \, g_{0,\ell+2}(x)
-  x \, g_{0,\ell+2}(\bx)  \over x - \bx }  = - u \,
{g_{1,\ell+1}(x) -  g_{1,\ell+1}(\bx)  \over x - \bx } \, . \cr}
}

For applications in the subsequent discussion the required expansions can in
general be reduced to considering the case $a=1$ when \defg\ satisfies
\eqn\gxp{
g_{1,j}(x) = (-1)^j g_{1,j}(x') \, \qquad x' = {x\over x-1} \, .
}
We then consider single variable expansions of the form
\eqn\Fexp{
F(x) = \sum_\ell a_\ell \, g_{1,\ell+1} (x) \, , 
}
where $F(x)$ is analytic in $x$ save for a branch cut along the real axis
from $1$ to $\infty$. The coefficients $a_\ell$ may be calculated by finding
a representation of the form
\eqn\Frep{
F(x) = \sum_{n=0}^N s_n x^{n+1} + \sum_{n=0}^{N'} s'{\!}_n x'^{n+1} +
\int_0^1 \d t \, r(t) \, {x \over 1- tx} \, ,
}
where $r(t)$ is determined by the discontinuity of $F$ across the branch cut. 
We may then note that
\eqn\pell{
x^{n+1} = - (-1)^n \sum_{\ell \ge n} {\ell!^2 \over (2\ell)!} \, {(\ell+n)!
\over n!^2 (\ell-n)!} \, g_{1,\ell+1}(x) \, ,
}
with a corresponding expansion for $x'^{n+1}$ as implied by \gxp.
Furthermore using\foot{Since ${\ell!^2 \over (2\ell)!}\, g_{1,\ell+1}(x)=(-1)^{\ell+1}
(\ell+{1\over 2})Q_\ell(z) $, where $z={2\over x}-1$ and $Q_\ell$ is an
associated Legendre function, this is equivalent to
${1\over z-y} = \sum_\ell (2\ell+1) P_\ell(y)Q_\ell(z)$.}
\eqn\expfa{
{ x \over 1-  t x} = - \sum_{\ell=0}^\infty {\ell!^2\over (2\ell)!} \,
(-1)^\ell \, P_\ell(2t -1) \,  g_{1,\ell+1}(x) \, , 
}
where $P_\ell$ are Legendre polynomials, in the integral involving $r(t)$ in \Fexp\ we 
may determine its contribution in terms of the integrals 
$\int_0^1 \d t \, r(t)\, P_\ell(2t-1)$ for $\ell=0,1,2,\dots$. By expanding in $t$ it
is easy to see that \expfa\ is equivalent to \pell.

\newsec{Superconformal Ward Identities and Operator Product Expansion}

The analysis of superconformal Ward identities is independent of $p$
so this is suppressed here.
In terms of the variables defined by \defxx\ and \defaa\ these require \NO
\eqn\Ward{
\G(u,v;\sigma,\tau )\big |_{\bar \alpha = {1\over \bar x}}
= F(x,\alpha) = k + \Big ( \alpha - {1\over x} \Big ) \hf ( x,\alpha) \,  .
}
Although not essential from the viewpoint of superconformal symmetry
dynamical requirements ensure that the function $f(x,\alpha)$ is identical
to the result obtained for free fields.
As a consequence of \crossG\ we have, with $x'$ as in \gxp,
\eqn\crossf{
\hf(x,\alpha) = - \hf(x',1-\alpha) \, .
}
The Ward identity \Ward\  may be solved by writing
\eqn\solW{\eqalign{
\G(u,v;\sigma,\tau) = {}& {k} + \G_\hf (u,v;\sigma,\tau) \cr
&{} + (\alpha x-1)(\alpha \bx-1)(\bar \alpha x - 1)(\bar \alpha \bx-1)
\H(u,v;\sigma,\tau)\, , \cr}
}
where the contribution from $\hf(x,\alpha)$ is given by
\eqn\Gfs
{\eqalign{
& \G_\hf (u,v;\sigma,\tau) + 2k  \cr
& = {(\bar \alpha x - 1)(\alpha \bx-1) 
\big ( F(x,\alpha)+F(\bx,\bar \alpha) \big ) - (\alpha x-1)(\bar \alpha \bx-1)
\big ( F(x,\bar\alpha)+F(\bx,\alpha) \big ) \over (x-\bx)(\alpha-\bar \alpha)}\, ,\cr}
}
with $F$ expressed in terms of $\hf$ as in \Ward, the contribution involving
$k$ cancels. Corresponding to
$\G^{(p)}$ then $\H^{(p)}(u,v;\sigma,\tau)$, as defined by \solW, is a 
polynomial in $\sigma, \,\tau$ of  degree $p-2$ and $\hf^{(p)}(x,\alpha)$ is 
a polynomial in $\alpha$ of degree $p-1$.
We may also note that
\eqn\as{\eqalign{
&(\alpha x-1)(\alpha \bx-1)(\bar \alpha x - 1)(\bar \alpha \bx-1) \cr
&{} = v + \sigma^2 uv + \tau^2 u + \sigma \, v(v-1-u) +
\tau (1-u-v) + \sigma \tau \, u(u-1-v) \, . \cr}
}
Non trivial dynamical contributions are only contained in
$\H(u,v;\sigma,\tau)$.

The representation \solW\ matches exactly with the results of the
operator product expansion and the requirements of superconformal symmetry.
The conformal partial wave expansion for $\G(u,v;\sigma,\tau)$ has in 
general the form
\eqn\expG{
\G(u,v;\sigma,\tau) = \sum_{nm,t_\ell \ell} a_{nm,t_\ell \ell} \, 
u^{t_\ell} G^{(\ell)}_{2t_\ell+\ell} (u,v) \, Y_{nm}(\sigma,\tau) \, ,
}
so that $a_{nm,t_\ell \ell}$ correspond to the contributions of operators 
belonging to the $SU(4)$ representation with Dynkin labels ${[n-m,2m,n-m]}$
and with spin $\ell$ and scale dimension $\Delta = 2t_\ell + \ell$.
For theories with superconformal symmetry the operators appearing in the
operator product expansion must belong to supermultiplets each of which
must have the same anomalous dimension. For $\N=4$ superconformal 
the supermultiplets are long $\A^\Delta_{nm,\ell}$, which may have an anomalous
dimension depending on the coupling, and semi-short $\C_{nm,\ell}$ or short
$\B_{nm}$, where the scale dimensions are protected against perturbative 
corrections.  For each possible $\N=4$ supermultiplet
which contributes to the operator product expansion the superconformal primary
operator with the lowest scale dimension $\Delta$ satisfies
\eqn\Dbound{
\Delta_{\rm BPS} = 2n \, , \qquad \Delta_{\rm semi} = 2n+2 + \ell \, , \qquad
\Delta_{\rm long} \ge 2n+2 + \ell \, .
}
There are also non unitary semi-short supermultiplets with 
$\Delta_{\rm semi} = 2m + \ell$. These play a role in the
superconformal decomposition of the operator product expansion \NO, but their
contributions are cancelled in unitary theories.
If a similar expansion to \expG\ for
$\H(u,v;\sigma,\tau)$,
\eqn\expH{
\H(u,v;\sigma,\tau) = \sum_{nm,j_\ell \ell} A_{nm,j_\ell \ell} \, 
u^{j_\ell} G^{(\ell)}_{2j_\ell+4+\ell}(u,v) \, Y_{nm}(\sigma,\tau) \, ,
}
is substituted in \solW, then for each term involving $A_{nm,j_\ell \ell}$
the factor \as\ generates a set of contributions for $a_{nm,t_\ell \ell}$
which correspond to the long supermultiplet $\A^{\Delta_\ell}_{nm,\ell}$ 
with the lowest state belonging to the representation $[n-m,2m,n-m]$ and spin
$\ell$ and scale dimension $\Delta_\ell = 2j_\ell + \ell$, 
all with the same anomalous dimension. 
Detailed expressions are contained in  \NO. Thus $\H$ determines the 
spectrum of all long multiplets which may contribute as conformal partial waves
in the operator product expansion for the four point function \fourp. For
$\H^{(p)}$ in \expH\  necessarily we must require $n\le p-2$.

For free theories, which are the starting point for perturbative treatments,
there are no anomalous dimensions and in \expG, \expH\ we may restrict
$t_\ell=t, \, j_\ell=j$ to be integers. It is convenient to first consider
an expansion in terms of single variable functions, following \cxexp,  so that
consequence we write
\eqn\Gexp{
(x-\bx) \G(u,v;\sigma,\tau)  =  x \sum_{t=0}^\infty 
\G_t(x,\sigma,\tau) \, g_{0,t}(\bx) \, , 
}
and then the complete partial wave expansion is obtained from
\eqn\Gcon{
\G_t(x,\sigma,\tau) = \sum_{\ell=-t-1}^\infty \sum_{nm} a_{nm,t\ell}
\, g_{1,t+\ell}(x) \, Y_{nm}(\sigma,\tau) \, .
}
In a similar fashion we write
\eqn\Hexp{
(x-\bx) \H(u,v;\sigma,\tau)  =  {1\over x} \sum_{j=0}^\infty
\H_j(x,\sigma,\tau) \, g_{2,j}(\bx) \, , 
}
with correspondingly,
\eqn\Hcon{
\H_j(x,\sigma,\tau) = \sum_{\ell=-j-1}^\infty \sum_{nm} A_{nm,j\ell}
\, g_{1,j+\ell+2}(x) \, Y_{nm}(\sigma,\tau) \, .
}
For consistency \crossG\ requires, with $x'$ as in \crossf, since
$g_{1,\ell}(x)= (-1)^\ell g_{1,\ell}(x')$,
\eqn\crossGH{
\G_t(x,\sigma,\tau) = (-1)^t \G_t(x',\tau,\sigma) \, , \qquad
\H_j(x,\sigma,\tau) = (-1)^j \H_j(x',\tau,\sigma) \, .
}

In general for any supermultiplet $\M$ we have a conformal partial wave 
expansion
\eqn\parM{
\G_{t}(\M;x,\sigma,\tau) = \sum_{nm,\ell} a_{nm,t\ell}(\M) \, g_{1,t+\ell}(x)
Y_{nm}(\sigma,\tau) \, ,
}
where the non zero $a_{nm,t\ell}(\M)$ 
correspond to the  spectrum of operators in the supermultiplet $\M$.
Any result for $a_{nm,t\ell}(\M)$  must be compatible with the 
relation from \rela\ $a_{nm,t\ell} = - a_{nm,t+\ell+1\, -\ell-2}$. 

Given the decomposition shown in \solW\ and \Gfs\ we may write
\eqn\GkfH{
\G_{t}(x,\sigma,\tau) = k\, \G_{t}(\I;x) + \G_{\hf,t}(x,\sigma,\tau)
+ \G_{\H,t}(x,\sigma,\tau) \, .
}
To obtain $\G_{\H,t}$ we use
\eqn\relg{
(\alpha \bx - 1) g_{a+1,j}(\bx) = - g_{a,j}(\bx) - (\alpha-\half)
g_{a,j+1}(\bx) - c_{j+a} g_{a,j+2}(\bx) \, ,
} 
where
\eqn\cj{
c_j = \gamma_{j,-1}\gamma_{j-1,1} = {j^2\over 4(2j-1)(2j+1)} \, ,
}
to obtain  $(\alpha \bx - 1)(\bar \alpha \bx - 1)  
g_{2,j}(\bx) = \sum_t c_{t,j}\, g_{0,t}(\bx)$ with
\eqn\ccc{\eqalign{
& c_{j,j} = 1 \, , \ \ c_{j+1,j} =  \half ( y + \bar y) \, , \ \
c_{j+2,j} = \quar\, y \bar y + c_j + c_{j+1}  \, , \cr
& c_{j+3,j} = c_{j+1} \half ( y + \bar y) \, , \quad 
c_{j+4,j} = c_{j+1} c_{j+2} \, ,\cr}
}
where $\half ( y + \bar y)= \sigma-\tau$ and $\quar\, y \bar y =
\half(\sigma+\tau-\half)$. With these results from \Gexp\ and \Hexp
\eqn\GHt{
\G_{\H,t}(x,\sigma,\tau) = 
\Big (\alpha -{1\over x}\Big )\Big (\bar \alpha - {1\over x} \Big )
\sum_{j\ge 0} c_{t,j} \, \H_j(x,\sigma,\tau) \, . 
}
The expansions \Gcon\ and \Hcon\ may then be related with the aid of
\eqn\recur{
\Big (\alpha -{1\over x}\Big )\Big (\bar \alpha - {1\over x} \Big )
g_{1,j+1}(x) = \sum_t c_{t,j} g_{1,t-1}(x) \, .
}

The contribution of a long multiplet $\A^{2j+k}_{nm,k}$ is defined
by letting in \Hcon\
$\H_j(x,\sigma,\tau) \to g_{1,j+\ell+2}(x) \, Y_{nm}(\sigma,\tau) $
so that \GHt\ and \parM\ give, using \recur,
\eqn\Long{
{\ts \sum_{rs}}\, a_{rs,t\ell}\big (\A^{2j+k}_{nm,k}\big )\, Y_{rs} = 
c_{t+\ell+1,j+k+1} \, c_{t,j}\,  Y_{nm} \, ,
}
where we must have $j\le t\le j+4$ and $j+k \le t+\ell \le j+k+4$. Detailed
expressions are easily obtained using \ccc\ and \recurY. Thus 
$\H_j(x,\sigma,\tau)$ determines the contribution of
all long multiplets whose lowest operator has twist $2j$.

In \GkfH\ $\G_t(\I)$ corresponds to the contribution for the unit operator 
and is determined by
\eqn\kt{
x-\bx = x \sum_{t=0}^1 \G_{t}(\I;x)\, g_{0,t}(\bx) \, ,
}
which gives
\eqn\Gkt{
\G_{0}(\I;x)  = 1 \, , \qquad \G_{1}(\I;x)= \half 
\Big ({1\over x} - {1\over x'} \Big ) = - g_{1,-1}(x) \, .
}
In this case we then have only the non zero contributions
\eqn\id{
a_{00,00}(\I) = - a_{00,1\,-2}(\I) = 1 \, .
}

To analyse the contributions arising from $\hf(x,\alpha)$, as 
expressed in $\G_{\hf,t}$, we use the expansion 
\eqn\expf{
\hf(x, \alpha)  
= \sum_{\ell=0}^\infty \hf_\ell(\alpha)  \, g_{1,\ell+1}(x) 
= \sum_{n,\ell} b_{n,\ell} \, P_n(y) \, g_{1,\ell+1}(x) \, ,
}
where $\hf_\ell(\alpha) = (-1)^\ell \hf_\ell(1-\alpha)$ and as shown later
$\hf_\ell(\alpha) = {\rm O}(\alpha^\ell)$, unless otherwise constrained by
the value of $p$. This ensures that $b_{n,\ell}$ is non zero only if
$n\le \ell,p-1$ and $n+\ell$ even. Using \Gexp\ for $\G_\hf$, as given by \Gfs,
with \expf\  then gives
\eqnn\Gft
$$\eqalignno{
\G_{\hf,t}(x,\sigma,\tau) = {}& 
- \Big (\alpha -{1\over x}\Big )\Big (\bar \alpha - {1\over x} \Big )
\bigg ( {\hf(x,\alpha) - \hf(x,\bar \alpha) \over \alpha - \bar \alpha}\, 
\de_{t\, 0} + {y \hf(x,\alpha) - {\bar y}\hf(x,\bar \alpha) \over 
2(\alpha - \bar \alpha)} \, \de_{t\, 1} \bigg ) \cr
&{} + \sum_{\ell \ge 0} c_{t,\ell} \,
{\big ( \alpha - {1\over x} \big ) \hf_\ell (\alpha) - 
\big ( \bar \alpha - {1\over x} \big ) \hf_\ell(\bar \alpha) \over 
\alpha - \bar \alpha} \, . & \Gft\cr}
$$
If we define 
\eqn\fz{
\hf(x,\alpha) = \hf_0(x,\alpha) + b_{0,0}\, g_{1,1}( x) \, ,
} 
and then with this decomposition \Gft\ can be expressed in the form
\eqn\defGf{
\G_{\hf,t}(x,\sigma,\tau) = b_{0,0}\, \G_{t}(\I;x) - {\ts{1\over 6}}b_{0,0}\,
\G_{t}(\B_{11};x,\sigma,\tau) + \G_{\hf_0,t}(x,\sigma,\tau) \, ,
}
where
\eqn\GBone{
{\ts{1\over 6}}\,
\G_{t}(\B_{11};x,\sigma,\tau) = \cases{\sum_{\ell=2}^4 c_{\ell,0} \, 
g_{1,\ell-1} (x) \, , &$t=1 \, ,$ \cr - c_{t,0} \, , &$t=2,3,4 \, .$\cr}
}
Using $c_{2,0}= {1\over 6}Y_{11}, \ c_{3,0}= {1\over 36}Y_{10}$ it is 
straightforward to see that in the conformal partial wave
expansion of $\G_{\B_{11}}$, as in \parM, this gives just the non zero 
results for $a_{nm,t\ell}(\B_{11})$,
\eqn\Bone{
a_{11,10}(\B_{11})  = 1 \, , \qquad
a_{10,11}(\B_{11})  = {\ts {1\over 6}} \, , \qquad
a_{00,12}(\B_{11})  = {\ts {1\over 30}} \, ,
}
apart from those related by the symmetry \rela.
This corresponds to the contribution of the lowest $\half$-BPS short
multiplet $\B_{11}$ containing the energy momentum tensor.

In a  free theory, where there are no anomalous dimensions, there is no direct
distinction between between the contribution of operators belonging to long
multiplet or semi-short multiplets which have protected scale dimensions. This
reflects the  the potential decomposition of the contribution of a long
multiplet at  the unitarity threshold into two corresponding semi-short
multiplets  as in \longm.  
Such ambiguities in the analysis of the operator product 
expansion must be resolved in a perturbative analysis of anomalous
dimensions.

To discuss the form of the contribution of semi-short multiplets
we first identify for a function $\F(x,\sigma,\tau)$ an expression
$\H^{(j)}_\F$, representing contributions with just twist $j$, which
is determined by
\eqn\Hj{\eqalign{
\H^{(j)}_{\F,k}(x,\sigma,\tau)  = {}& \de_{kj}\,  \F(x,\sigma,\tau) 
- \F_k(\sigma,\tau) \, g_{1,j+1}(x) \, ,   \cr
\F(x,\sigma,\tau) = {}& \sum_n \F_n(\sigma,\tau)\, g_{1,n+1}(x) \, . \cr}
}
Here we must have $\F(x,\sigma,\tau) = (-1)^j \F(x',\tau,\sigma)$.
Correspondingly we define
\eqn\Gt{
\G^{(j)}_{\F,t}(x,\sigma,\tau)  =  \de_{tj} \,
\Big (\alpha-{1\over x}\Big )\Big (\bar \alpha-{1\over x}\Big )
\F(x,\sigma,\tau) - \sum_{k=0}^t c_{t,k} \, \F_k(\sigma,\tau) \, 
g_{1,j-1}(x) \, . 
}
The right hand sides of \Hj\ and \Gt\ are such that in the expansions
\Hcon\ and \Gcon\ they are compatible with the conditions
$A_{nm, j \ell} = - A_{nm,j+\ell+1\, - \ell-2}$
and $a_{nm, t \ell} = - a_{nm,t +\ell+1\, - \ell-2}$, as required by \rela.

We then define for the function $\hf(x,\alpha)$
\eqn\QQ{\eqalign{
\Q^0_j(x,\sigma,\tau) 
= {}& - {P_j(\bar y) \, \hf(x,\alpha) - P_j( y) \, \hf(x,\bar \alpha)
\over \alpha - \bar \alpha} \, , \cr
\Q^1_j(x,\sigma,\tau) = {}& - 
{P_j(\bar y) \, y\hf(x,\alpha) - P_j( y) \, \bar y 
\hf(x,\bar \alpha) \over 2(\alpha - \bar \alpha)} \, , \cr}
}
More generally the contribution of semi-short multiplets are determined 
by $\hf(x,\alpha)$ and expressed in terms of $\hG^{(j)}_{\hf,t}$
which is of the form, using the notation in \Leg\ and \Gt, 
\eqn\Gshort{\eqalign{
\hG^{(j)}_{\hf,t}(x,\sigma,\tau) = {}& \G^{(j)}_{\Q^0_j,t}(x,\sigma,\tau)
+ \G^{(j+1)}_{\Q^1_j,t}(x,\sigma,\tau) + \gamma_{j,-1}\,
\G^{(j+1)}_{\Q^0_{j-1},t}(x,\sigma,\tau) \cr
&{}+ \gamma_{j,-1} \, \G^{(j+2)}_{\Q^1_{j-1},t}(x,\sigma,\tau) 
+ c_j \, \G^{(j+2)}_{\Q^0_j,t}(x,\sigma,\tau) \cr
&{} + \gamma_{j,-1}\, c_{j+1} \,
\G^{(j+3)}_{\Q^0_{j-1},t}(x,\sigma,\tau) \, .\cr}
}
Using \Gft\ we may easily recognise that
\eqn\Gsem{
\G_{\hf,t}(x,\sigma,\tau) = \G^{(0)}_{\Q^0_0,t}(x,\sigma,\tau)
+ \G^{(1)}_{\Q^1_0,t}(x,\sigma,\tau) = \hG^{(0)}_{\hf,t}(x,\sigma,\tau) \, .
}

If we denote by $\C_{nm,\ell}$ a semi-short multiplet in which the lowest
state belongs to the $SU(4)$ representation $[n-m,2m,n-m]$ and has spin $\ell$,
so that $\Delta - \ell = 2n+2$, then the contributions to the partial wave
expansion are determined by
\eqn\Gsemi{
\hG^{(j)}_{\hf,t}(x,\sigma,\tau) \Big |_{\displaystyle{ \hf (x,\alpha) =
\half  P_i(y) g_{1,j+k+2}(x)}}=  
\G_t\big (\C_{{j-1}\, i, k};x,\sigma,\tau \big ) \, ,
}
for $j\ge i+1, \ k \ge 0$. In this case \QQ\ gives $\Q^0_j = Y_{j-1\, i}\,
g_{1,j+k+2}$ so that it is easy to see that the contribution of the lowest
dimension operator in \parM\ is given by 
$a_{j-1\,i,j k} (\C_{{j-1}\, i, k} ) = 1$.
The remaining contributions to $a_{nm,t\ell} (\C_{{j-1}\, i, k})$, with
$t=j, \dots , j+3$ and $\ell= k-3, \dots, k+4 $,
may be easily calculated from \Gshort\ and \Gt\ using \recurY. For
application to the conformal partial wave expansion we may note that \Gsemi\
can be extended by using,
\eqn\ashort{
\G_t\big (\C_{{j-1}\, i, -1} \big ) = \cases{\gamma_{j,1}\, 
\G_t (\B_{ji}) \, , &$j > i\, $,\cr
\noalign{\vskip 2pt}
\gamma_{j,1} \big ( \G_t (\B_{jj} ) -
\gamma_{j+1,1}\gamma_{j,1} \, \G_t (\B_{j+1\,j+1})\big )\, ,& $j=i\,$,\cr}
}
where the first case follows from \shortm. As in the introduction
$\B_{ji}$, $j\ge i$ denotes a short multiplet whose lowest
state belongs to the representation $[j-i,2i,j-i]$ with $\Delta = 2j$. For
$j>i$ this is a $\quar$-BPS multiplet whereas $\B_{jj}$ is a $\half$-BPS
multiplet. The decomposition of $\G_t(\B_{ji})$ for 
$i=0,\dots, j$, which determines the contributions for the short BPS 
multiplets, is exhibited in appendix A.

Furthermore using the relations \Leg\ we may combine two contributions of
the form \Gshort\ to form those for a long multiplet
\eqn\ssl{\eqalign{
\hG^{(j)}_{\hf,t}(x,\sigma,\tau) +  \gamma_{j,1}  \, 
\hG^{(j+1)}_{\hf,t}(x,\sigma,\tau) = {}& 
\Big (\alpha -{1\over x}\Big )\Big (\bar \alpha - {1\over x} \Big ) 
\sum_k c_{t,k} \, \H^{(j)}_{\Q^0_j,k}(x,\sigma,\tau) \cr
={}& \G_{\H,t}(x,\sigma,\tau) \ \ \hbox{for} \ \ \H_k = \H^{(j)}_{\Q^0_j,k}
\, , \cr}
}
using \GHt\ and in accord with \Hj,
\eqn\HQk{\eqalign{
\H^{(j)}_{\Q^0_j,k}(x,\sigma,\tau)= {}& \de_{k j}\,  \Q^0_j(x,\sigma,\tau) - 
\Q^0_{j,k}(\sigma,\tau)\, g_{1,j+1}(x) \, , \cr
\Q^0_{j,k}(\sigma,\tau) = {}& - {P_j(\bar y) \, \hf_k(\alpha) - P_j( y) \, 
\hf_k(\bar \alpha) \over \alpha - \bar \alpha} \, . \cr}
}
The relation \ssl\ is a reflection of the decomposition \longm. Using \ssl\ the 
complete result
\eqn\GsH{
\G_{t}(x,\sigma,\tau) = k\, \G_{t}(\I;x) + 
\hG^{(0)}_{\hf,t}(x,\sigma,\tau) + \G_{\H,t}(x,\sigma,\tau) \, .
}
may be rewritten by progressively removing contributions involving
$\hG^{(j)}_{\hf,t}(x,\sigma,\tau)$ for $j=0,1,\dots$ at the expense of modifying 
$\H$. This reflects the inherent ambiguity in a free theory in the
decomposition into contributions from long and semi-short operators.
By letting $\H \to \hH$ which is given by
\eqn\HH{
\hH_k(x,\sigma,\tau) = \H_k(x,\sigma,\tau) + \sum_{j=0}^{J-1} (-1)^j
{j!^2\over (2j)!} \, \big ( \de_{k j}\,  \Q^0_j(x,\sigma,\tau) - 
\Q^0_{j,k}(\sigma,\tau)\, g_{1,j+1}(x) \big ) \, ,
}
then instead of \GsH\ we have
\eqn\GJH{ 
\G_{t}(x,\sigma,\tau)  = k\, \G_{t}(\I;x) + (-1)^J{J!^2\over (2J)!} \,  
\hG^{(J)}_{\hf,t}(x,\sigma,\tau) + \G_{\hH,t}(x,\sigma,\tau) \, . 
}
This removes the contribution of any semi-short multiplets with twist
$\Delta-\ell  < 2J $.

However to determine the full decomposition into supermultiplets it
is necessary to separate all contributions from short multiplets.
To achieve this we define, extending \fz,
\eqn\fj{
\hf(x,\alpha) = \hf_j(x,\alpha) + \sum_{\ell=0}^j f_\ell(\alpha) 
g_{1,\ell+1}(x) \, , \quad f_\ell(\alpha) = 
\sum_{n=0}^\ell b_{n,\ell} \, P_n(y) \, , \quad j < p \, .
}
{}From \ashort\ we may then obtain
\eqn\GBB{
\hG^{(j)}_{\hf_{j-1},t} = \hG^{(j)}_{\hf_j,t} + 2\gamma_{j,1} \Big ( 
\sum_{i=0}^{j-1} b_{i,j} \, \G_t(\B_{ji}) + b_{j,j}
\big ( \G_t(\B_{jj}) - \gamma_{j,1}\gamma_{j+1,1}\, 
\G_t(\B_{j+1\,j+1}) \big ) \Big ) \, .
}
Using \GBB\ there is a then corresponding modification of \ssl\  which 
allows short multiplet contributions to be explicitly identified,
\eqnn\sslj
$$\eqalignno{
& \hG^{(j)}_{\hf_j,t}(x,\sigma,\tau) +  \gamma_{j,1}  \,
\hG^{(j+1)}_{\hf_{j+1},t}(x,\sigma,\tau) \cr
\noalign{\vskip -2pt}
&{} =  - 2\gamma_{j,1} \gamma_{j+1,1} \Big ( 
\sum_{i=0}^{j} b_{i,j+1} \, \G_t(\B_{j+1\,i}) + b_{j+1,j+1}
\big ( \G_t(\B_{j+1\,j+1}) - \gamma_{j+1,1}\gamma_{j+2,1}\,
\G_t(\B_{j+2\,j+2}) \big ) \Big )  \cr
\noalign{\vskip -4pt}
&\quad {}+  \Big (\alpha -{1\over x}\Big )\Big (\bar \alpha - {1\over x} \Big )
\sum_k c_{t,k} \big ( \de_{k j}\,  \hat \Q^0_j(x,\sigma,\tau) - 
\hat \Q^0_{j,k}(\sigma,\tau)\, g_{1,j+1}(x) \big ) \, , & \sslj \cr}
$$
where since $\hf \to \hf_j$ we have $\hat \Q^0_j = \Q^0_j - \sum_{k=0}^j \Q^0_{j,k}$,
$\hat \Q^0_{j,k}=0 $ for $k \le j$.
Applying this result we now have
\eqn\GJHB{\eqalign{
\G_{t} = {}& (k + b_{0,0})\, \G_{t}(\I) + 
2 \sum_{j=1}^{J-1} (-1)^j {(j+1)!^2 \over (2j+2)!} \, \big (
b_{j,j} + \gamma_{j-1,1} \, b_{j-1,j-1} \big ) \G_t(\B_{jj}) \cr
&{} + (-1)^J {(J+1)!^2\over (2J+2)!} \, 
{J \over 2J-1} \, b_{J-1,J-1} \, \G_t(\B_{JJ}) \cr
&{} + 2 \sum_{j=1}^{J} (-1)^j {(j+1)!^2 \over (2j+2)!} \sum_{i=0}^{j-1} 
b_{i,j} \, \G_t(\B_{ji}) + (-1)^J{J!^2\over (2J)!} \, \hG^{(J)}_{\hf_J,t} 
+ \G_{\hH,t} \, . \cr}
}
Here the definition of $\hH$ is modified from \HH\ by $\Q^0_j \to \hat \Q^0_j$
and using $\hf_J(x,\alpha) = \sum_{m,\ell\ge 0} b_{m,J+\ell+1} P_m(y)
g_{1,J+\ell+2}(x)$ with \Gsemi\ we have
\eqn\GC{
\hG^{(J)}_{\hf_J,t} = 2 \sum_{m,\ell\ge 0} b_{m,J+\ell+1} \,
\G_t \big (\C_{J-1\, m, \ell} \big ) \, .
}
For application to the conformal partial decomposition of chiral four
point functions in the large $N$ limit we identify $J=p$. Since in \GC\
$m=0,\dots,p-1$ this expansion with $J=p$ gives the contribution of 
semi-short multiplets in the operator product expansion, unitarity requires 
$(-1)^p b_{m,\ell} \ge 0$ for $\ell>p$.
All contributions appearing in \GJHB\ are then identifiable with various
possible supermultiplets.

\newsec{Free Field Results}

In order to discuss the dynamical contributions it is first necessary to
obtain the results for the free part of the general chiral four point function.
The free field contribution satisfies the Ward identities \Ward\ by
requiring it to have the form
\eqn\Free{
\G^{(p)}(u,v;\sigma,\tau) = \sum_{r,s\ge 0 \atop r+s \le p} a_{rs} \, 
(\sigma u)^r \Big ( \tau {u\over v} \Big )^s \, .
}
{}From \Ward\ we then have
\eqn\freef{
F^{(p)}(x,\alpha) = \sum_{r,s} a_{rs} \, 
(\alpha x)^r \big ( (1-\alpha) x'\big )^s \, , \qquad k = \sum_{r,s} a_{rs} \, .
}
However for the four point function \fourp\ crossing symmetry \crossG\ provides 
further crucial constraints on the
coefficients $a_{rs}$. This requires that \Free\ may be rewritten in terms 
of two variable crossing symmetric polynomials.

Crossing symmetric polynomials in the variables 
$\sigma,\tau$ of degree $n$, i.e they may be expanded in monomials 
$\sigma^g \tau^h$ with $g+h\le n$, are defined by
\eqn\csr{
S^{(n)}(\sigma,\tau)=S^{(n)}(\tau,\sigma)=\tau^n S^{(n)}(\sigma/\tau,1/\tau)\, .
}
A simple basis for these polynomials is given by 
\eqn\csb{
S^{(n)}_{ab}(\sigma,\tau)=\cases{\sigma^a \tau^a + \sigma^a \tau^{n-2a}
+ \sigma^{n-2a} \tau^a , &$a=b\,$,\cr 
\sigma^a \tau^b + \sigma^b \tau^a + \sigma^a \tau^a , &$2a+b=n$,\cr
\sigma^{{1\over 3}n} \tau^{{1\over 3}n} , &${\ts{1\over 3}}n\in {\Bbb N}$,\cr
\sigma^a \tau^b \! + \sigma^b \tau^a \! 
+ \sigma^a \tau^{n-a-b} \! + \sigma^{n-a-b} \tau^a \! + \sigma^b \tau^{n-a-b}
\! + \sigma^{n-a-b} \tau^b , &otherwise,\cr}
}
where $a,b$ are integers satisfying
\eqn\ab{
0 \le b \le a \, , \qquad 2a+b \le n \, .
}
The first two cases in \csb\ are distinguished according to whether $n>3a$
or $3a>n$ respectively.
For any $n$  the set of possible $(a,b)$ are the points of an integer lattice 
inside or on a triangle with vertices $(0,0)$, $({1\over 3}n, {1\over 3}n)$
and $({1\over 2}n,0)$.
The number of independent such crossing symmetric polynomials was derived in
\refs{\Howe,\NO}, their results are equivalent to $[{1\over 12}n(n+6)]+1$. From \csb\
it is trivial to verify that the number of terms $n^{(n)}_{ab}$ in the above 
symmetric polynomials is in each case
\eqn\sab{
n^{(n)}_{aa\vphantom b}  = n^{(n)}_{a\, n-2a} = 3 \, , \quad
n^{(n)}_{{1\over 3}n \, {1\over 3}n} = 1 \, , \quad
n^{(n)}_{ab} = 6 \ \ \hbox{otherwise} \, ,
}
and then
\eqn\check{
\sum_{{0\le b \le a \atop \le n-a-b}} n^{(n)}_{ab} = \half (n+1)(n+2) \, ,
}
which is the number of independent polynomials in $\sigma,\tau$ of degree $n$.

We may now satisfy \crossG\ by writing instead of \Free
\eqn\FreeC{
\G^{(p)}(u,v;\sigma,\tau) = S^{(p)}_{00} (\sigma u , \tau u/v) 
+{p^2 \over N^2} \sum_{{0\le b \le a \atop \le p-a-b}} w^{(p)}_{ab}
S^{(p)}_{ab} (\sigma u , \tau u/v) \, , 
}
where $w^{(p)}_{00}=0$ and we have imposed the normalisation condition
$\G^{(p)}(0,v;\sigma,\tau)=1$. In general the $w^{(p)}_{ab}$ are complicated
functions of $N$ but for single trace operators in the large $N$ limit, as shown in
appendix B, we have
\eqn\wN{
w^{(p)}_{a0} = 1 \, , \qquad w^{(p)}_{ab} = 2 \, , \quad b\ge 1 \, .
}
This ensures that \FreeC\ may be written for any $p$ in the form
\eqn\FreeN{\eqalign{
\G^{(p)}(u,v;\sigma,\tau) = {}& 1
+{p^2 \over N^2}  \sum_{r=1}^{p-1} u^r \bigg ( \sigma^r  + 
\tau^r {1\over v^r} + 2 \sum_{s = 1}^{r-1}
\sigma^{r-s} \tau^s {1 \over v^s} \bigg ) \cr
\noalign{\vskip -4pt}
&{} +  u^p \bigg ( \sigma^p + \tau^p {1\over v^p} + {p^2 \over N^2}
\sum_{r=1}^{p-1} \sigma^{p-r}\tau^r {1\over v^r} \bigg )  \, , \cr}
}
where the terms involving $\sigma^{r-s} \tau^s u^{r}$ with $0<r<p$ are 
universal in that they are independent of $p$ apart from the overall 
coefficient $p^2/N^2$.

{}From \FreeN\ and \Ward\ we easily obtain
\eqn\freek{
k = 3 + {p^2\over N^2}(p^2-1) \, ,
}
and 
\eqn\freep{
\G^{(p)}(u,v;\sigma,\tau) = 1 + {p^2\over N^2} \, {\tilde \G}(u,v;\sigma,\tau)
+ {\rm O}(u^p) \, ,
}
where
\eqn\defGt{
{\tilde \G}(u,v;\sigma,\tau) = 
{u \sigma + u\tau/v  \over (1-u\sigma)(1-u\tau/v)} \, ,
}
is a universal function, independent of $p$.  We also then have
\eqn\freef{
\hf^{(p)}(x,\alpha) = \bigg ( k-1 + {p^2 \over N^2} \Big ( \alpha(1-\alpha)
{\d^2 \over \d \alpha^2} + (1-2\alpha) {\d \over \d \alpha} \Big ) \bigg )
{ x \over 1-\alpha x} + {\rm O}(x^{p+1}) \, .
}
Using \expfa\ we easily obtain from  \expf
\eqn\fell{
\hf_\ell(\alpha) = (-1)^\ell {\ell!^2\over (2\ell)!} \Big ( 1- k
+ {p^2 \over N^2} \, \ell (\ell+1) \Big ) P_\ell (y) \, , \qquad
\ell = 0, \dots p-1 \, .
}
It is then clear that $b_{n,\ell}$ in \expf\ is non zero only for $n=\ell$ if 
$\ell<p$. For application in \GJHB\ we may note that the contribution of 
$\half$-BPS multiplets is given by
\eqn\sss{
(-1)^j \big (b_{j,j} + \gamma_{j-1,1} \, b_{j-1,j-1}\big ) = {j!^2 \over (2j-1)!} \, 
{p^2 \over N^2} \, , \qquad j =1, \dots , p-1 \, .
}
In general as a consequence of \Gsemi\ and \ashort\ $\{b_{n,\ell}\}$ determines
for which cases the index $I_{nm}$ defined in \ind\ is non zero.

We here adapt the discussion in the previous section to determine the
free field form for ${\hat \H}_t$,  which was defined in \HH\ and is responsible 
for the contributions of long multiplets. First with the definition \QQ\ it is 
convenient to further define, following \Gshort,
\eqn\FQ{\eqalign{
\F^{(j)}_t = {}& \de_{t\, j} \Q^0_j + \de_{t\, j+1} \big ( \Q^1_j + \gamma_{j,-1}
\Q^0_{j-1} \big ) \cr
&{} + \de_{t\, j+2} \big ( \gamma_{j,-1} \Q^1_{j-1} + c_j \, \Q^0_j \big )
+ \de_{t\, j+3} \, \gamma_{j,-1} c_{j+1}\, \Q^0_{j-1} \, . \cr}
}
This satisfies
\eqn\FFH{
\F^{(j)}_t(x,\sigma,\tau) + \gamma_{j,1} \, \F^{(j+1)}_t(x,\sigma,\tau) 
= \sum_j c_{t,j} \, \Q^0_j (x,\sigma,\tau) \, ,
}
which is essentially equivalent to \ssl. Starting from \GkfH\ with \GHt\ and 
\Gft\ then
\eqn\decGH{\eqalign{ \!\!\!\!
\G^{(p)}_{t}& (x,\sigma,\tau) - k\, \G_{t}(\I;x) -
\sum_{\ell \ge 0} c_{t,\ell} \,
{\big ( \alpha - {1\over x} \big ) \hf_\ell (\alpha) -
\big ( \bar \alpha - {1\over x} \big ) \hf_\ell(\bar \alpha) \over
\alpha - \bar \alpha} \cr \!\!\!\!
&{}= \Big ( \alpha -{1\over x} \Big ) \Big ( \bar \alpha -{1\over x} \Big )
\bigg ( \sum_j c_{t,j} \,  \H_j (x,\sigma,\tau)  + \F^{(0)}_t (x,\sigma,\tau) \bigg ) \cr
&{}= \Big ( \alpha -{1\over x} \Big ) \Big ( \bar \alpha -{1\over x} \Big )
\bigg ( \sum_j c_{t,j} \,  \hH_j (x,\sigma,\tau)  + 
(-1)^J {J!^2 \over (2J)!} \, \F^{(J)}_t (x,\sigma,\tau) \bigg ) \, . \cr}
}
where in this case we have by using \FFH
\eqn\HHn{
\hH_j (x,\sigma,\tau) = \H_j (x,\sigma,\tau) +
{(-1)^j}{j!^2 \over (2j)!} \, \Q^0_j (x,\sigma,\tau) \, , \quad
j = 0,1, \dots , J-1 \, .
}
This definition of $\hH_j$ is essentially equivalent to \HH, using \HHn\
the expansion $\hH_j (x,\sigma,\tau) = \sum_\ell A_{j\ell}(\sigma,\tau)
g_{1,j+\ell+2}(x)$ determines the corresponding expansion for $\hH_j$ given 
by \HH\ with coefficients $\half( A_{j\, \ell} - A_{j+\ell+1\ -\ell-2})$, 
manifestly satisfying \rela. The result \decGH\ is similarly equivalent to \GJH.
Following previous considerations we take $J=p$. 

The result \decGH\ may be simplified by noting that
\eqn\ident{
\sum_{\ell \ge 0} c_{t,\ell} \, (-1)^\ell {\ell!^2 \over (2\ell)!} \,
{\big ( \alpha - {1\over x} \big ) P_\ell (y)  -
\big ( \bar \alpha - {1\over x} \big ) P_\ell(\bar y) \over
\alpha - \bar \alpha} = \G_{t}(\I;x) \, ,
}
since then the $1-k$ terms in \fell\ are just sufficient to change the
coefficient of $\G_{t}(\I;x)$ in \decGH\ to $1$. From \freep\ we have
\eqn\Gtt{
\G^{(p)}_t (x,\sigma,\tau) = \G_{t}(\I;x) + {p^2 \over N^2} \, 
{\tilde \G}_t (x,\sigma,\tau)  \, , \qquad t < p \, ,
}
and hence \decGH, with \fell, gives
\eqn\univ{\eqalign{
{\tilde \G}_t (x,\sigma,\tau) = {}& \sum_{\ell > 0} c_{t,\ell} \, 
(-1)^\ell {\ell!^2 \over (2\ell)!} \, \ell(\ell+1) \, 
{\big ( \alpha - {1\over x} \big ) P_\ell (y)  -
\big ( \bar \alpha - {1\over x} \big ) P_\ell(\bar y) \over
\alpha - \bar \alpha} \cr
&{}+ \Big ( \alpha -{1\over x} \Big ) \Big ( \bar \alpha -{1\over x} \Big )
\sum_j c_{t,j} \,  \K_j (x,\sigma,\tau) \, , \cr}
}
where
\eqn\HHt{
\hH_j (x,\sigma,\tau) = {p^2 \over N^2 } \, 
\K_j (x,\sigma,\tau) \, , \qquad j < p \, .
}
It is clear that $\K_j$ is universal, independent of the
particular $p$. 

The left hand side of \univ\ is determined by the expansion
\eqn\expGt{
\Big ({1\over x}- {1\over \bx} \Big )\, {\tilde \G}(u,v;\sigma,\tau) = 
\sum_t {\tilde \G}_t (x,\sigma,\tau) \, g_{1,t-1}(\bx) \, .
}
Writing  from \defGt
\eqn\Gexp{\eqalign{
\Big ({1\over x}- {1\over \bx} \Big ) {\tilde \G}(u,v;\sigma,\tau) =
- \sigma x + \tau x'&{} +  \sigma \, {1- \sigma x + \tau x' \over
1 - \sigma x - \tau x'}\, (1- \sigma x^2) \, { \bx \over 1 - \sigma x\, \bx}\cr
&{}+ \tau \, {1+ \sigma x - \tau x' \over 1 - \sigma x - \tau x'} 
\, (1- \tau x'^2) \, { \bx \over 1 - (1- \tau x')  \bx}\, ,
\cr}
}
we have
\eqn\Gone{
{\tilde \G}_1(x,\sigma,\tau) = - \sigma x + \tau x' \, ,
}
and also, using \expfa, then ${\tilde \G}_t$ is determined for $t=2,3,\dots$
by
\eqn\Gtt{\eqalign{
{\tilde \G}_{t+2}(x,\sigma,\tau) = - {t!^2 \over (2t)!} \bigg ( &
\sigma \, {1- \sigma x + \tau x' \over
1 - \sigma x - \tau x'}\, (1- \sigma x^2) \, P_t ( 1 - 2\sigma x) \cr
\noalign{\vskip -1pt}
&{}+ \tau \, {1+ \sigma x - \tau x' \over 1 - \sigma x - \tau x'}
\, (1- \tau x'^2) \, P_t ( 2 \tau x' - 1 ) \bigg ) \, . \cr}
}
It is easy to see that this satisfies \crossGH. For application in \univ\
we consider modified expansion
\eqn\expGct{
\Big ({1\over x}- {1\over \bx} \Big )\, {\tilde \G}(u,v;\sigma,\tau) =
\sum_{t,j} c_{t,j}\, {\tilde \G}'{\!\!}_j (x,\sigma,\tau) \, g_{1,t-1}(\bx) \, .
}
Using \recur\ this is equivalent to
\eqn\expGtt{
{\bx^2 \over (1 - \alpha \bx)(1- {\bar \alpha} \bx)}  \,
\Big ({1\over x}- {1\over \bx} \Big )\, {\tilde \G}(u,v;\sigma,\tau) =
\sum_{j}  {\tilde \G}'{\!\!}_j (x,\sigma,\tau) \, g_{1,j+1}(\bx) \, .
}
By decomposing into partial fractions of the form \expfa\ we obtain
\eqn\Gj{\eqalign{
{\tilde \G}'{\!\!}_j (x,\sigma,\tau) = {}& - (-1)^j {j!^2 \over (2j)!} \,
{(1-x)^2 \over (1 - \alpha x)^2(1- {\bar \alpha} x)^2} \cr
\noalign{\vskip -2pt}
& \ {} \times \Big ( 
b \, P_j ( 2\sigma x - 1 ) + c \, P_j ( 1- 2 \tau x' ) +
{ a \, P_j(y) - {\bar a} \, P_j ({\bar y}) \over \alpha-{\bar \alpha}} \Big ) 
\, , \cr
b = {}& {(1- \sigma x^2)(1- \sigma x + \tau x') \over 1-x} \, , \quad
c = (1-x) (1- \tau x'^2)(1 +  \sigma x - \tau x') \, , \cr
a = {}& {(1 - \alpha x)^3 ( {\bar y} - {\bar \alpha} x) \over (1-x)^2} \, ,
\qquad \qquad \!
{\bar a} = {(1 - {\bar \alpha} x)^3 ( {y} - {\alpha} x) \over (1-x)^2} \, .
\cr}
}
This then gives
\eqn\Hjr{
\K_j (x,\sigma,\tau) =  - (-1)^j {j!^2 \over (2j)!} \,
\Big ( B P_j ( 2\sigma x - 1 ) + C \, P_j ( 1- 2 \tau x' )
+ {A_j \, P_j(y) - {\bar A}_j \, P_j ({\bar y}) \over \alpha-{\bar \alpha}}
\Big ) \, ,
}
with 
\eqn\BA{\eqalign{
B= {}& - 2{\bar \alpha}x^2 \, {(1-\alpha)(1-{\bar\alpha})\over
(1-{\bar\alpha}x)^3(\alpha-{\bar\alpha})^3}(\alpha -{\bar\alpha}^2 x)
- {{\bar \alpha}x^2 \over (1-{\bar\alpha}x)^2(\alpha-{\bar\alpha})}
+ \alpha \leftrightarrow {\bar\alpha} \, , \cr
A_j = {}& { x^2 ( {\bar y} - {\bar \alpha} x) \over (1-{\bar\alpha}x)^3} 
- j(j+1) \, { x \over 1-{\bar\alpha}x } \, , \quad
{\bar A}_j = A_j \big |_{\alpha \leftrightarrow {\bar\alpha}} \, , \quad
C = B \big |_{\alpha\to 1-\alpha, {\bar\alpha} \to 1 - {\bar\alpha},x\to x'}
\, .  \cr}
}

In general the results given by \Hjr\ with \BA\ are not straightforward
to analyse. However for the leading term for large $\sigma,\tau$  we have
\eqn\Hlead{\eqalign{
\K_j (x,\sigma,\tau) \sim (-1)^j \bigg (& \big ( \sigma (\sigma x)^j - \tau
(- \tau x')^j - (\sigma-\tau)^{j+1} \big ) {\sigma x - \tau x' \over
(\sigma x + \tau x')^3} \, x^2 x'^2 \cr
\noalign{\vskip - 4pt}
{}& - j(j+1)\, (\sigma - \tau)^j \, {xx' \over \sigma x + \tau x'} \bigg ) \, .
\cr}
}
Hence writing in general
\eqn\Hcexp{
\K_j (x,\sigma,\tau) = (-1)^j \sum_{g,h\ge 0} \K_{gh,j}(x) \, 
\sigma^g \tau^h \, , \quad \K_{gh,j}(x) = (-1)^j \K_{hg,j}(x') \, ,
}
then $\K_{gh,j}= 0 $ for $g+h \ge j$ and from expanding \Hlead\ so that
the denominators $\sigma x + \tau x'$ are cancelled,
\eqnn\cexp
$$\eqalignno{
\K_{gh,g+h+1}(x) = (-1)^h \bigg ( & (h+1)^2 x^{g+1} x'^{h+2} - x'^2
\sum_{r=0}^h {g+h+2 \choose r} (h+1-r)^2 \Big ( {x'\over x} \Big )^{h-r}\cr
\noalign{\vskip - 4pt}
&{} - x' (g+h+1)(g+h+2) 
\sum_{r=0}^h {g+h+1 \choose r} \Big ( {x'\over x} \Big )^{h-r}\bigg )\, . & \cexp \cr}
$$
Using $x^nx'^m = {1\over (m-1)!} \sum_{r=1}^n \! {(n+m-r-1)!\over (n-r)!} x^r
+ {1\over (n-1)!} \sum_{r=1}^m \! {(n+m-r-1)!\over (m-r)!} x'^r$ for 
$n,m\ge 1$ and $x'/x=x'-1$ this may be written as a polynomial in $x$ of degree
$g+1$ and in $x'$ of degree $h+1$. For general $j$ there is no apparent
general formula for $\K_{gh,j}(x)$ but we note that
\eqn\Kzero{
\K_{00,j}(x) = {j!^2 \over (2j)!} \, j(j+1) \big ( x + (-1)^j x'\big ) \, ,
}
and we have also obtained $\K_{10,j}(x)$ in accord with results in appendix C.

\newsec{Expressions for General Chiral Four Point Functions}

In this section we suggest  general expressions for the dynamical part of
the large $N$ amplitude  for the four point function \fourp\  for identical single 
trace $\half$-BPS operators belonging to the $SU(4)$ $[0,p,0]$ representation.
This will be based on what has been observed in the examples for 
$p=2,3,4$ which have been explicitly calculated using the AdS/CFT 
correspondence \refs{\AFd,\ADHS,\Degen}. For $p\ge 4$ there are also multi-trace 
BPS operators but our considerations do not apply for these.

As described in section 4 the dynamical part of the four point function, for 
general $p$, reduces to $\H^{(p)}$, polynomial of degree $p-2$ in $\sigma,\tau$, 
which as a consequence of \crossG\ satisfies the crossing relations
\eqn\csh{
\H^{(p)}(u,v;\sigma,\tau)={1\over v^2}\, \H^{(p)}(u/v,1/v;\tau,\sigma)=
\left({u\over v}\right)^{p-2}\! \tau^{p-2} \, \H^{(p)}(v,u;\sigma/\tau,1/\tau).
}
The results for $p=2,3,4$ can be reduced to the following form
\eqn\lna{
\H^{(p)}(u,v;\sigma,\tau)=-{p^2\over N^2} \ u^p  \!\!
\sum_{{0\le b \le a \atop 2a+b \le p-2}}
\sum_{i,j,k} \, c^{(p)}_{ijk,ab}\, T^{(p)}_{ijk,ab}(u,v;\sigma,\tau) \, , 
}
where $T^{(p)}_{ijk,ab}$ are completely crossing symmetric combinations of $\oD$ 
functions which are related to the crossing symmetric polynomials in \csb. For
$p=2,3,\dots$ and restricting $0 \le b \le a , \ 2a+b \le p-2$, corresponding to
\ab, we define
\eqn\T{\eqalign{ \!\!\!
T^{(p)}_{ijk,ab}(u,v;\sigma,\tau) & \cr
= {\ts{1\over 6}}n^{(p-2)}_{ab}&  \big (
\sigma^a\tau^b\, \oD_{i\,p+2\,jk}(u,v)+ \sigma^b\tau^a\, \oD_{i\, p+2\,kj}(u,v)\cr
&{} + \sigma^a\tau^{p-2-a-b}\, \oD_{j\, p+2\, ik}(u,v)+
\sigma^{p-2-a-b}\tau^a\, \oD_{j\, p+2\, ki}(u,v) \cr
&{}+ \sigma^b\tau^{p-2-a-b}\, \oD_{k\, p+2\, ij}(u,v)+
\sigma^{p-2-a-b}\tau^b\, \oD_{k\, p+2\, ji}(u,v) \big ) \, . \cr}
}
The crossing identities for $\oD$ functions ensure that 
$u^p T^{(p)}_{ijk,ab}(u,v;\sigma,\tau)$ satisfies \csh\ without any additional
factors of $u,v$ being necessary. The factors ${\ts{1\over 6}}n^{(p-2)}_{ab}$
are introduced for later convenience, essentially since, for the boundary 
values of $a,b$, we have
\eqn\symT{\eqalign{
& T^{(p)}_{ijk,aa} = T^{(p)}_{ikj,aa} \, , \qquad
T^{(p)}_{ijk,a\,p-2-2a} = T^{(p)}_{kji,a\, p-2-2a} = T^{(p)}_{jki,aa} \, , \cr
& T^{(p)}_{ijk,{1\over3}(p-2)\,{1\over3}(p-2)} = 
T^{(p)}_{(ijk),{1\over3}(p-2)\,{1\over3}(p-2)} \, . \cr}
}
When $a,b$ satisfy the conditions in \symT\ then \T\ can be simplified in
appropriate cases,
\eqn\TT{\eqalign{
&{} T^{(p)}_{ijj,aa}(u,v;\sigma,\tau)= T^{(p)}_{jij,a\, p-2-2a}(u,v;\sigma,\tau)\cr
&{} = \sigma^a\tau^a\, \oD_{i\,p+2\,jj}(u,v)+ \sigma^a\tau^{p-2-2a}\, 
\oD_{j\, p+2\, ij}(u,v) 
+ \sigma^{p-2-2a} \tau^a\, \oD_{j\, p+2\, ji}(u,v)  \, , \cr
&{} T^{(p)}_{iii,{1\over3}(p-2)\,{1\over3}(p-2)} (u,v;\sigma,\tau)
= (\sigma \tau)^{{1\over3}(p-2)} \oD_{i\,p+2\,ii}(u,v) \, . \cr }
}

Further restrictions arise from the the unitarity conditions for each 
long multiplet which contribute in the operator product expansion. For
the lowest scale dimension operator belonging to the representation
$[n-m,2m,n-m]$ and with spin $\ell$ we must have $\Delta \ge 2n+\ell+2$. 
To apply this if we expand
\eqn\exH{
\H^{(p)}(u,v;\sigma,\tau) = \sum_{0\le m \le n\le p-2} A_{nm}(u,v)
\, Y_{nm}(\sigma,\tau) \, ,
}
where $Y_{nm}(\sigma,\tau)$ are the polynomials in \poly, then
the unitarity condition requires
\eqn\unit{
A_{nm}(u,v)= {\rm O}(u^{n+1})  \, .
}

The $\oD$ functions that can appear in \T\ are constrained by the unitarity 
conditions \unit. To obtain these we first list the essential properties used 
here, see \refs{\DO,\ADHS}. For $\oD_{n_1n_2n_3n_4}(u,v)$ we define
\eqn\dss{
\Sigma = \half ( n_1 + n_2 + n_3 + n_4) \, , \qquad
s = \half ( n_1 + n_2 - n_3 - n_4) \, ,
}
and, restricting to $\Sigma$ an integer, for $s=1,2,\dots$ we have
\eqnn\Df
$$\eqalignno{
\oD_{n_1n_2n_3n_4}(u,v) = {}& \sum_{m=0}^{s-1} u^{-s+m} 
{(-1)^m \over m!} \, (s-m-1)! \,f_{n_1-s+m\, n_2-s+m\,n_3+m\, n_4+m} (v)\cr 
&{} + \ln u \, a_{n_1n_2n_3n_4}(u,v) + b_{n_1n_2n_3n_4}(u,v) \, ,  & \Df\cr}
$$
where $a_{n_1n_2n_3n_4}(u,v), \, b_{n_1n_2n_3n_4}(u,v)$ are both given by power 
series in $u,1-v$. For $s=0$ the first term in \Df\ is omitted. 
In \Df\ the relevant terms here are
given in terms of ordinary hypergeometric functions
\eqn\fm{
f_{n_1 n_2 n_3 n_4} (v) = 
{\Gamma(n_1)\Gamma(n_2)\Gamma(n_3)\Gamma(n_4) \over 
\Gamma(n_3+n_4)}\, F(n_2,n_3;n_3+n_4;1-v) \, .
}
{}From standard relations for hypergeometric functions we have
\eqn\hyper{
n_1 f_{n_1n_2n_3n_4}(v)= f_{n_1+1\, n_2\, n_3+1\, n_4}(v)
+ f_{n_1+1\, n_2n_3\,n_4+1}(v) \, ,
}
and also
\eqn\crossf{
f_{n_1 n_2 n_3 n_4} (v) = v^{-n_2}f_{n_1 n_2 n_4 n_3} (1/v) \, .
}
The $\ln u$ terms in \Df\ result in anomalous dimensions for operators
belonging to long multiplets which have twist $\Delta - \ell \ge 2p$ at zeroth
order in the $1/N$ expansion, where $\Delta$ is the scale dimension and $\ell$
the spin.  The terms involving negative powers $u^{-s+m}$ have no corresponding 
$\ln u$ terms and would correspond in the operator product expansion to 
contributions from long multiplets which are unrenormalised. We assume
that these are all cancelled although there remain contributions from various
semi-short multiplets which cannot be combined to form a long multiplet, thus
all multiplets not satisfying a shortening condition gain anomalous dimensions
in the $1/N$ expansion as expected.
This is the essential assumption that leads to strong constraints on the 
expansion \lna.

Using the properties of the $SU(4)$ harmonics $Y_{nm}(\sigma,\tau)$ it is
clear that the unitarity condition \unit\ is satisfied if
for any contribution in \lna\ involving
\eqn\ssD{
\sigma^g \tau^h \, u^p \oD_{n_1 \, p+2 \, n_3 n_4}(u,v) \, ,
}
we require
\eqn\gh{
0 \le s \le p - g - h -1 \, .
}
With the assumed expansion given by \lna\ and \T\
the condition $s\ge 0$ ensures from \Df\ that only long multiplets
with twist $\Delta - \ell \ge 2p$ with non zero  anomalous dimensions in the 
large $N$ limit can contribute to the operator product expansion of the
chiral four point function.

Applying the condition \gh\ to \T\ gives the following inequalities
\eqn\ies{\eqalign{
p-2a\leq {}& i+j-k\leq p+2\, ,\cr
p-2b\leq {}& i+k-j\leq p+2\, ,\cr
2(a+b+2)-p \leq {}& j+k-i\leq p+2 \, .\cr}
}
We also impose
\eqn\ca{
i,j,k\leq p \, .
}
It is clear that there is a finite number of possibilities for $i,j,k$, note
that $p+4 \le i+j+k \le 3p$.
We should also note that the expansion \T\ is not unique since
\eqn\DDD{
\half( i+j+k - p -2) \oD_{i\, p+2\, j\,k+1} =
\oD_{i\, p+2\, j+1\,k+1} + \oD_{i+1\, p+2\, j\,k+1} + \oD_{i+1\, p+2\, j+1\,k}\,.
}
This is the only relation for $\oD$ functions of the form appearing in \T.
Correspondingly we may take
\eqn\TTT{
\half( i+j+k - p -2) T^{(p)}_{ijk,ab} = T^{(p)}_{i\,j+1\,k+1,ab} +
T^{(p)}_{i+1\,j\,k+1,ab}  + T^{(p)}_{i+1\,j+1\, k,ab} \, ,
}
which allows the expansion \lna\ to be simplified if each term in \TTT\ 
satisfies the constraints \ies\ and \ca. For this to be the case
we must have
\eqn\ijk{
i+j-k, i+k-j, j+k-i \le p \, , \qquad i,j,k \le p-1 \, .
}
Whenever \ijk\ is satisfied one of the terms appearing in \TTT\ may be omitted
in the general expansion.

If we use \ijk\ to remove terms with the lowest value of $i+j+k$  whenever
appropriate then for general $p$ the list of possible terms obtained from 
\ies, \ca\ modulo \ijk\ is
\eqn\pp{\eqalign{
&T^{(p)}_{pjj,ab}\, ,\ \quad j=a+b+2,\dots , p \, , \cr
&T^{(p)}_{iip,ab}\, ,\ \quad i=p-a,\dots , p-1 \, , \  a \ge 1 \, , \cr
&T^{(p)}_{ipi,ab}\, ,\ \quad i=p-b,\dots , p-1 \, , \  b \ge 1 \, , \cr
&T^{(p)}_{i \, i+k -p-2 \, k,ab}\, ,\ \ i=p-a+1,\dots , p \, , \ 
k=a+b+3,\dots , p \, , \ a \ge 1 \, , \cr
&T^{(p)}_{i j\, i+j -p-2 ,ab}\, ,\ \ \, i=p-b+1,\dots , p \, , \
j=a+b+3,\dots , p \, , \ b \ge 1 \, , \cr
&T^{(p)}_{j+k -p-2 \, j k,ab}\, ,\ \ j=p-a+1,\dots , p \, , \
k=p-b+1 ,\dots , p \, , \ a,b \ge 1 \, . \cr}
}
When $a=b$ or $b=p-2-2a$ it is necessary also to take into account the
symmetry conditions in \symT\ to obtain an independent basis.
If $N^{(p)}_{ab}$ are the number of possibilities for each $a,b$ then we have
\eqn\num{\eqalign{
N^{(p)}_{ab} = {}& (a+b+1)(p-a-b-1) + ab \, , \quad
0 \le b< a < \half ( p-2-b) \, , \cr
N^{(p)}_{aa} = {}&  (a+1)(p-{\ts {3\over 2}}a - 1) \, , \qquad
0 \le a < {\ts {1\over 3}} (p-2)  \, , \cr
N^{(p)}_{a\, p-2-2a} = {}& (a+1)(p-{\ts {3\over 2}}a -1) \, , \qquad
{\ts {1\over 3}} (p-2) < a \le \half ( p-2) \, , \cr
N^{(p)}_{{1\over 3}(p-2) \, {1\over 3}(p-2)} = {}& 
{\ts {1\over 18}}(p+1)(p+4) \, , \ \quad \qquad p=2 \ \hbox{mod} \ 3 \, . \cr}
}

The crucial assumption, in addition to \T, is that the leading terms in the
expansion of $\H^{(p)}$ in powers of $u$, which do not involve any $\ln u$ 
terms, are universal, i.e. we have for any $p=2,3, \dots$
\eqn\Huniv{
\H^{(p)}(u,v;\sigma,\tau)=-{p^2\over N^2} \, \F(u,v;\sigma,\tau) 
+ {\rm O}(u^{p}) \, ,
}
where $\F$ is independent of $p$. From \csh\ we have
\eqn\crossF{
\F(u,v;\sigma,\tau) = {1\over v^2} \, \F(u/v,1/v;\tau,\sigma) \, ,
}
and it is expressible in general as an expansion
\eqn\expFF{
\F(u,v;\sigma,\tau) = \sum_{n\ge g+h+2}\sum_{g,h\ge 0} 
\sigma^g \tau^h \, u^{n-1} \F_{n,gh}(v) \, .
}
With the expansion \lna\ for $\H^{(p)}$, and with $T^{(p)}_{ijk,ab}$  given
by \T, then only the leading singular terms in the expansion of the $\oD$
functions shown in \Df\ contribute to $\F$ in \Huniv. Assuming
$T^{(p)}_{ijk,ab}$ are restricted as in \pp\ then the potential contributions
to $\F_{n,gh}$ in \expFF\ which are given by \Huniv\ are just
\eqn\listf{\eqalign{
&{} f_{n-1\,n+1\, nn} \, , \qquad n= g+h+2, \dots , p \, ,\cr
&{} f_{j-1\,n+1\, nj} \, , \qquad j= n-h, \dots , n \, , \ n=p \, , \, ,\cr
&{} f_{j-1\,n+1\, jn} \, , \qquad j= n-g, \dots , n \, , \ n=p \, , \, ,\cr
&{} f_{j\,n+1\, j\, n+1} \, , \qquad j= n-g, \dots , n-1 \, , \ 
n=g+h+2, \dots p \, , \cr
&{} f_{j\,n+1\, n+1\, j} \, , \qquad j= n-h, \dots , n-1 \, , \
n=g+h+2, \dots p \, , \cr}
}
which may be further reduced with the aid of \hyper.

A additional restriction, which is compatible with the results for $p=2,3,4$,
is that only the terms with $n=g+h+2$ are present in \expFF\ so that
\eqn\expF{\eqalign{
\F(u,v;\sigma,\tau) = {}& \sum_{g,h\ge 0} 
\sigma^g \tau^h \, u^{g+h+1}  \F_{gh}(v) \cr
\F_{gh}(v) = {}& \sum_{l,m} d_{gh,lm} \, 
f_{l+m-g-h-3\ g+h+3 \ lm }(v) \, , \cr}
}
where $l,m$ are restricted in accord with \listf\  with $n=g+h+2$.
Apart from the relation \hyper\ the functions $u^{n} f_{l+m-n-2\, n+2 \,lm}(v)$
appearing in \expF\ are assumed to be linearly independent. With the restricted
form in \expF\ then, as shown later, we are both able to determine $\F$ from
\Huniv\ and using the explicit form of $\F$, up to terms of ${\rm O}(u^{p-1})$, 
to find all the coefficients $c^{(p)}_{ijk,ab}$  which appear in the expansion 
\lna.

A simple consequence of \Huniv\ and \lna, which does not require any 
restrictions of the form for $\F$, is that for $p=3,4,\dots$,
\eqn\univ{\eqalign{
u^p \!\!\! & \sum_{{0\le b \le a \atop 2a+b \le p-2}}
\sum_{i,j,k} \, c^{(p)}_{ijk,ab}\, T^{(p)}_{ijk,ab}(u,v;\sigma,\tau)\cr
&{} = u^{p-1} \!\!\! \sum_{{0\le b \le a \atop 2a+b \le p-3}}
\sum_{i,j,k} \, c^{(p-1)}_{ijk,ab}\, T^{(p-1)}_{ijk,ab}(u,v;\sigma,\tau)
+ {\rm O}(u^{p-1}) \, , \cr}
}
where only the leading singular terms displayed explicitly in \Df\ need be 
considered.  These equations are invariant under $\sigma \leftrightarrow \tau$
and $u\to u/v, \, v \to 1/v$.
In \univ\ all $\oD$ functions with $s=1,2, \dots$ are relevant on 
the right hand side but only those with $s=2,3,\dots$ on the left hand side.
It is important to note that \univ\ does not constrain,
$c^{(p)}_{ppp,ab}$, the coefficient for $\oD_{p\,p+2\,pp}$ which is present
for any $a,b$.

\newsec{Applications for Low $p$}
We now show how the above suggestions work out in practice for low $p$, initially
using only \univ.

For $p=2$ $\H^{(2)}$ is independent of $\sigma,\tau$ and there is just one
possible $\oD$ function which is in accord with
the simplification of results obtained from AdS/CFT \refs{\AFd,\SCFT},
\eqn\Htwo{
\H^{(2)}(u,v)= -{4\over N^2}\, u^2\oD_{2422}(u,v) \, .
}
For $p=3$ we must have again $a=b=0$ and there are just two crossing symmetric 
forms, \ADHS, 
\eqn\Hthree{\eqalign{
\H^{(3)}& (u,v;\sigma,\tau) =-{9\over N^2}u^3 \Big (
c^{(3)}_{322,00}T^{(3)}_{322,00}+c^{(3)}_{333,00}T^{(3)}_{333,00} \Big ) \, ,\cr
={}&-{9\over N^2}\, u^3 \Big (c^{(3)}_{333,00} (1+\sigma+\tau)\oD_{3533}+
c^{(3)}_{322,00}\big (\oD_{3522}+\sigma\oD_{2523}+\tau\oD_{2532}\big )\Big )\, .
\cr}
}
The equations \univ\ give just one condition arising from
$u^3\oD_{3522}(u,v) \sim u f_{1322}(v)$ and in \Htwo\
$u^2\oD_{2422}(u,v) \sim u f_{1322}(v)$ so that we require
\eqn\cth{
c^{(3)}_{322,00} = 1 \, .
}
The known results also give
\eqn\cp{
c^{(3)}_{333,00} = 1 \, .
}

For $p=4$, for comparison with previous results, we write
\eqn\Hfour{\eqalign{
\H^{(4)}(u,v;\sigma,\tau)={}&
-{16\over N^2}\, u^4\Big (c^{(4)}_{422,00}T^{(4)}_{422,00}+
c^{(4)}_{433,00}T^{(4)}_{433,00}+c^{(4)}_{444,00}T^{(4)}_{444,00}\cr&
\qquad\qquad+c^{(4)}_{323,10}T^{(4)}_{323,10}+c^{(4)}_{424,10}T^{(4)}_{424,10}
+c^{(4)}_{444,10}T^{(4)}_{444,10} \Big )\, ,\cr}
}
where from \symT\ and \TTT\ $T^{(4)}_{334,10} = T^{(4)}_{433,10} = \half
(T^{(4)}_{323,10} - T^{(4)}_{424,10})$ so that such contributions are discarded.
Expanding \Hfour\ then gives
\eqn\Hexpr{\eqalign{
\H^{(4)}(u,v;\sigma,\tau)
= -{16\over N^2}\, u^4 \Big (&  
\big (c^{(4)}_{444,00}(1+\sigma^2+\tau^2)+c^{(4)}_{444,10}
(\sigma+\tau+ \sigma\tau)\big )\oD_{4644}\cr
&{} + c^{(4)}_{323,10}
\big (\sigma\oD_{3623}+\tau\oD_{3632}+\sigma\tau\oD_{2633}\big )\cr
&{} + c^{(4)}_{424,10}
\big (\sigma\oD_{4624}+\tau\oD_{4642}+\sigma\tau\oD_{2644}\big )\cr
&{} +c^{(4)}_{422,00}\big (\oD_{4622}+\sigma^2\oD_{2624}+\tau^2\oD_{2642}\big )\cr
&{} +c^{(4)}_{433,00}\big (\oD_{4633}+\sigma^2\oD_{3634}+\tau^2\oD_{3643}\big )
\Big ) \, . \cr}
}
Applying \univ\ then gives for the 1 terms
\eqn\one{
c^{(4)}_{422,00} = \half \, , \qquad c^{(4)}_{422,00} - c^{(4)}_{433,00} = 0 \, ,
}
and from the $\sigma$ terms
\eqn\sig{
c^{(4)}_{323,10}\, f_{1423}(v) +  c^{(4)}_{424,10}\, f_{2424}(v) =
f_{2433}(v) +  f_{1423}(v) \, .
}
Using \hyper\ this is easily solved giving
\eqn\cfour{
c^{(4)}_{323,10} =  2 \, , \qquad c^{(4)}_{424,10} = - 1 \, .
}
The remaining coefficients which are undetermined by \univ\ are
\eqn\cfourp{
c^{(4)}_{444,10} =  1 \, , \qquad c^{(4)}_{444,00} =  {\ts {1\over 4}}  \, .
}
For the basis corresponding to \pp\ then, instead of \Hexpr, we should take
\eqn\Hexpt{\eqalign{
\H^{(4)}(u,v;\sigma,\tau)
= -{16\over N^2}\, u^4 \Big ( & 
{\ts {\sum_{j=2}^4}} c^{(4)}_{4jj,00} T^{(4)}_{4jj,00} \cr
&{} + 2 c^{(4)}_{433,10} T^{(4)}_{433,10} + 
c^{(4)}_{424,10} T^{(4)}_{424,10} + c^{(4)}_{444,10} T^{(4)}_{444,10}  \Big ) \, .
\cr}
}
Here we introduce a factor 2 for the $c^{(4)}_{433,10}$ terms to count equal 
contributions from $T^{(4)}_{433,10}$ and $T^{(4)}_{343,10}$. This ensures
uniformity with later general results. 
In this case \one\ and \cfour\ are unchanged but instead of \cfour\ we should
take
\eqn\cfoura{
c^{(4)}_{433,10} =  2 \, , \qquad c^{(4)}_{424,10} = 1 \, .
}

On the basis of the results for $p=2,3,4$ we determine the first few terms in 
the function $\F$ introduced in \Huniv,
\eqn\Funiv{\eqalign{
\F(u,v;\sigma,\tau) = {}& u f_{1322}(v) \cr
&{}+ \sigma \, u^2 \big ( f_{1423}(v) + f_{2433}(v) \big ) 
+ \tau \,  u^2 \big ( f_{1432} (v) + f_{2433}(v) \big ) \cr
&{} + \sigma^2 u^3 \half \big ( f_{1524}(v) + f_{2534}(v) + \half f_{3544}(v)\big )\cr
&{} + \tau^2 u^3 \half \big ( f_{1542}(v) + f_{2543}(v) + \half f_{3544}(v) \big )\cr
&{}+ \sigma \tau \, u^3  \big ( 2f_{1533}(v) + f_{3544}(v) \big ) + {\rm O}(u^4)\, .  \cr}
}
This result is in accord with the assumed form in \expF.

For $p=5$ we have the general form
\eqn\Hfive{\eqalign{
\H^{(5)}(u,v;\sigma,\tau)
=  -{25\over N^2}\, u^5 \Big ( & {\ts \sum_{j=2}^5} 
c^{(5)}_{5jj,00}T^{(5)}_{5jj,00}\cr&
{} + {\ts \sum_{j=3}^5} c^{(5)}_{5jj,10}T^{(5)}_{5jj,10} 
+ c^{(5)}_{445,10}T^{(5)}_{445,10} \cr&
{} +c^{(5)}_{524,10}T^{(5)}_{524,10}
+ c^{(5)}_{535,10}T^{(5)}_{535,10}\cr&
{}+ 3 c^{(5)}_{544,11}T^{(5)}_{544,11} +
3 c^{(5)}_{553,11}T^{(5)}_{553,11}+ c^{(5)}_{555,11}T^{(5)}_{555,11}\Big )\, .\cr}
}
Here we note from \symT\ that $T^{(5)}_{ijk,11}$ is completely symmetric
in $i,j,k$ and ${1\over 3}T^{(5)}_{333,11} = T^{(5)}_{443,11} = 
T^{(5)}_{544,11} + \half T^{(5)}_{553,11}$, so neither of these terms are present
in the expansion in \Hfive. By using \TTT, we also
eliminate terms involving $T^{(5)}_{434,10}, \, T^{(5)}_{423,10}$. As in 
\Hexpt\ we introduce factors to take account of the sum over identical terms
related by \symT.
With the basis in \Hfive\ and the results for $\H^{(4)}$ we may readily
solve the equations \univ\ giving
\eqn\resfive{\eqalign{
& c^{(5)}_{522,00} = c^{(5)}_{533,00} = {\ts {1\over 6}} \, , \qquad
c^{(5)}_{544,00} = {\ts {1\over 12}} \, , \cr
& c^{(5)}_{533,10} = c^{(5)}_{544,10} = 1 \, , \qquad
c^{(5)}_{524,10} = c^{(5)}_{535,10} = \half \, , \qquad
c^{(5)}_{445,10} = {\ts {3\over 4}} \, , \cr
& c^{(5)}_{544,11} = 3 \, , \qquad c^{(5)}_{553,11} = 1 \, . \cr}
}
Only $c^{(5)}_{555,00},c^{(5)}_{555,10},c^{(5)}_{555,11}$ are undetermined at 
this stage. In the expansion \expFF\ we may now obtain
\eqn\Ffour{\eqalign{
\sum_{gh} \sigma^g\tau^h \F_{5,gh}(v) = {}& 
\big ( c^{(5)}_{555,00} - {\ts {1\over 36}} \big) f_{4655}(v) \cr 
\noalign{\vskip -8pt}
& {}+ (\sigma+\tau+\sigma^2+\tau^2) 
\big ( c^{(5)}_{555,10}- \quar \big )f_{4655}(v) \cr
&{}+ \sigma \tau \big ( c^{(5)}_{555,11}-  1 \big )f_{4655}(v)  \cr
&{}+ \sigma^3 \big ( {\ts {1\over 6}} f_{1625}(v) + {\ts {1\over 6}} f_{2635}(v)
+ {\ts {1\over 12}} f_{3645}(v) + c^{(5)}_{555,00}  f_{4655}(v) \big ) \cr
&{}+ \tau^3 \big ( {\ts {1\over 6}} f_{1652}(v) + {\ts {1\over 6}} f_{2653}(v)
+ {\ts {1\over 12}} f_{3654}(v) + c^{(5)}_{555,00}  f_{4655}(v) \big ) \cr
&{}+ \sigma^2\tau \big ( f_{1634}(v) + \half f_{2644}(v)
+ \quar f_{3645}(v) + c^{(5)}_{555,10}  f_{4655}(v) \big ) \cr
&{}+ \sigma\tau^2 \big ( f_{1643}(v) + \half f_{2644}(v)
+ \quar f_{3654}(v) + c^{(5)}_{555,10}  f_{4655}(v) \big ) \, . \cr}
}
The restrictions, imposed by the additional constraint that just the leading 
term for $n=g+h+1$ appears in the expansion of $\F$ for each $g,h$ as assumed in
\expF, is easily achieved in \Ffour\ by setting
\eqn\solfour{
c_{555,00}^{(5)} = {\ts {1\over 36}} \, , \qquad
c_{555,10}^{(5)} = \quar \, , \qquad c_{555,11}^{(5)} = 1 \, .
}

For $p=6$ we have the following expansion in terms of independent crossing 
symmetric functions, noting that $T^{(6)}_{ijk,11}=T^{(6)}_{ikj,11}$
and $T^{(6)}_{635,20}=T^{(6)}_{536,20}$,
\eqn\Hsix{\eqalign{
\H^{(6)}(u,v;\sigma,\tau)
=  -{36\over N^2}\, u^6 \Big ( & {\ts \sum_{j=2}^6}
c^{(6)}_{6jj,00}T^{(6)}_{6jj,00}\cr&
{} + {\ts \sum_{j=3}^6} c^{(6)}_{6jj,10}T^{(5)}_{6jj,10}
+ c^{(6)}_{556,10}T^{(6)}_{556,10} \cr&
{} + c^{(6)}_{624,10}T^{(6)}_{624,10}
+ c^{(6)}_{635,10}T^{(6)}_{635,10} +  c^{(6)}_{636,10}T^{(6)}_{636,10} \cr&
{}+ {\ts \sum_{j=4}^6} c^{(6)}_{6jj,11}T^{(6)}_{6jj,11} +
2c^{(6)}_{556,11}T^{(6)}_{556,11} \cr
& {} + 2c^{(6)}_{635,11}T^{(6)}_{635,11}
+ 2c^{(6)}_{646,11}T^{(6)}_{646,11} +  c^{(6)}_{466,11}T^{(6)}_{466,11} \cr
&{} + 2c^{(6)}_{635,20}T^{(6)}_{635,20}
+ c^{(6)}_{646,20}T^{(6)}_{646,20} +  c^{(6)}_{525,20}T^{(6)}_{525,20} \Big )\, .\cr}
}
Here for $a=b=1$ and $a=2,b=0$ we have used \symT\ to reduce the number
of necessary terms. In this case the equations give
\eqn\csix{\eqalign{
& c^{(6)}_{622,00}= c^{(6)}_{633,00} = {\ts {1\over 24}} \, , \quad
c^{(6)}_{644,00} = {\ts {1\over 48}}  \, , \quad 
c^{(6)}_{655,00} = - {\ts {1\over 48}} + c^{(5)}_{555,00}  \, , \cr
& c^{(6)}_{633,10}= c^{(6)}_{644,10} = {\ts {1\over 3}} \, , \quad
c^{(6)}_{655,10} = - {\ts {1\over 12}} + c^{(5)}_{555,10} \, , \cr
& c^{(6)}_{624,10}= c^{(6)}_{635,10} = {\ts {1\over 6}} \, , \quad
c^{(6)}_{646,10}= {\ts {1\over 12}} \, ,
\quad  c^{(6)}_{556,10}= {\ts {1\over 12}} + c^{(5)}_{555,00} \, , \cr
& c^{(6)}_{644,11}= {\ts {3\over 2}} \, , \quad c^{(6)}_{655,11} =
{\ts {1\over 2}} + c^{(5)}_{555,11} \, , \quad 
c^{(6)}_{466,11} = {\ts {1\over 4}} \, , \cr
& c^{(6)}_{635,11}= c^{(6)}_{646,11} = {\ts {1\over 2}} \, , \quad
c^{(6)}_{556,11}= {\ts {3\over 4}} + c^{(5)}_{555,10} \, , \cr
& c^{(6)}_{644,20}=  {\ts {3\over 8}} \, , \quad 
c^{(6)}_{655,20}= {\ts {1\over 8}} + c^{(5)}_{555,10} \,
\quad c^{(6)}_{525,20}= c^{(6)}_{646,20} = {\ts {1\over 4}} \, , \quad
c^{(6)}_{635,20} = {\ts {1\over 4}} \, . \cr }
}

We may now extend the results given by \Funiv\ and \Ffour, assuming \solfour, to
obtain
\eqnn\Ffive
$$\eqalignno{
\sum_{gh} \sigma^g\tau^h \F_{6,gh}(v) = {}&
\big ( c^{(6)}_{666,00} - {\ts {1\over 576}} \big) f_{5766}(v) \cr 
\noalign{\vskip -8pt}
&{}+ (\sigma+\tau+\sigma^3+\tau^3) \big ( c^{(6)}_{666,10}- 
{\ts {1\over 36}}  \big )f_{5766}(v) \cr
&{}+ \sigma \tau (1+\sigma +\tau) \big ( c^{(6)}_{666,11}-  \quar \big )
f_{5766}(v)  \cr
&{}+ (\sigma^2 +\tau^2) \big ( c^{(6)}_{666,20}-  {\ts{1\over 16}} \big )
f_{5766}(v)  \cr  
&{}+ \sigma^4 \big ( {\ts {1\over 24}} f_{1726}(v) + {\ts {1\over 24}}
f_{2736}(v) + {\ts {1\over 48}} f_{3746}(v) + {\ts {1\over 144}} f_{4756}(v)
+ c^{(6)}_{666,00} f_{5766}(v) \big ) \cr
&{}+ \tau^4 \big ( {\ts {1\over 24}} f_{1762}(v) + {\ts {1\over 24}}
f_{2763}(v) + {\ts {1\over 48}} f_{3764}(v) + {\ts {1\over 144}} f_{4765}(v)
+ c^{(6)}_{666,00} f_{5766}(v) \big )
\cr &{}+ \sigma^3\tau \big ( {\ts {1\over 3}} f_{1735}(v) + {\ts {1\over 3}}
f_{2745}(v) + {\ts {1\over 18}} f_{4756}(v) + c^{(6)}_{666,10}f_{5766}(v) \big ) \cr
&{}+ \sigma\tau^3 \big ( {\ts {1\over 3}} f_{1753}(v) + {\ts {1\over 3}}
f_{2754}(v) + {\ts {1\over 18}} f_{4765}(v) + c^{(6)}_{666,10}f_{5766}(v) \big ) \cr
&{}+ \sigma^2\tau^2 \big ( {\ts {3\over 4}} f_{1744}(v) + {\ts {3\over 8}}
f_{3755}(v) + c^{(6)}_{666,20}  f_{5766}(v) \big ) \, . & \Ffive \cr}
$$
Again the same fashion as \solfour\ and in accord with \expF\ we also obtain
\eqn\solfive{
c_{666,00}^{(6)} = {\ts {1\over 576}} \, , \qquad
c_{666,10}^{(6)} = {\ts {1\over 36}} \, , \qquad c_{666,11}^{(6)} = 
{\ts {1\over 4}} \, , \qquad c_{666,20}^{(6)} = {\ts {1\over 16}} \, . 
}

\newsec{General Solutions}

We here discuss how the equations which follow from \Huniv, assuming
\lna\ with $T^{(p)}_{ijk,ab}$ restricted as in  \pp, can be solved if
we also suppose that the only contributions in the expansion for $\F$ 
are restricted to the form shown in \expF. The general expansion has the
form
\eqnn\Hexpall
$$\eqalignno{
\H^{(p)}=-{p^2\over N^2} \ u^p \!\!\!\!\sum_{{0\le b \le a \atop 2a+b \le p-2}} 
\!\!\! \bigg ( & 
\sum_{j=a+b+2}^p \! c^{(p)}_{pjj,ab} \, T^{(p)}_{pjj,ab} +
\sum_{i=p-a}^{p-1}\, c^{(p)}_{iip,ab} \, T^{(p)}_{iip,ab} +
\sum_{i=p-b}^{p-1}\, c^{(p)}_{ipi,ab} \, T^{(p)}_{ipi,ab} \cr
&{} + \sum_{i=p-a+1}^p \sum_{k=a+b+3}^p c^{(p)}_{\,i\, i+k-p-2 \, k,ab} \,
T^{(p)}_{i\, i+k-p-2 \, k,ab} \cr
&{} + \sum_{i=p-b+1}^p \sum_{j=a+b+3}^p c^{(p)}_{\,i j\, i+j-p-2,ab} \, 
T ^{(p)}_{i\, j\, i+j-p-2,ab} \cr
& {}+ \sum_{j=p-a+1}^p \sum_{k=p-b+1}^p \, c^{(p)}_{j+k-p-2\, jk,ab}\, 
T^{(p)}_{j+k-p-2\, jk,ab} \bigg ) \, . & \Hexpall \cr}
$$

We first consider terms independent of $\sigma,\tau$ which arise only from
$\sum_{j=2}^p c^{(p)}_{pjj,00} T^{(p)}_{pjj,00}$ where $T^{(p)}_{pjj,00} \to
\oD_{p\, p+2 \, jj}$. Requiring 
$\F(u,v;\sigma,\tau) = u f_{1322}(v) + {\rm O}(\sigma,\tau)$ as in \Funiv\ we 
obtain
\eqn\abz{
\sum_{j=2}^p \sum_{m=0}^{p-j} {(-1)^m\over m!}(p-j-1)! \, 
c^{(p)}_{pjj,00} u^{j+m-1} f_{j-1+m\, j+1+m\, j+m \, j+m}(v) = 
u f_{1322}(v) \, ,
}
which requires
\eqn\solz{
(p-2)! \, c^{(p)}_{p22,00} = 1 \, , \qquad 
\sum_{m=0}^{k-2} {(-1)^m\over m!} \, c^{(p)}_{p\, k-m\,k-m,00} = 0 \, , \  \
k =3, \dots , p \, .
}
This is easily solved giving
\eqn\solcz{
c^{(p)}_{pjj,00} = {1\over (p-2)!\, (j-2)!} \, , \qquad j=2,\dots, p \, .
}

We next consider the calculation of the coefficients $c^{(p)}_{ijk,ab}$ for 
$a\ge 1, \, b=0$. These are determined in \Huniv\ by the terms in the 
expansion \expF\ with  $g=a, \, h=0$. Contributions proportional to 
$\sigma^g$ first arise in an expansion in powers of $u$ of $\H^{(p)}$ 
from $\sum_{j=2}^p \, c^{(p)}_{pjj,00}\, T^{(p)}_{pjj,00}$,
with $T^{(p)}_{pjj,00}(u,v;\sigma,\tau) \to \sigma^p \oD_{j\,p+2\,jp}(u,v)$,
for $p=g+2$.
Using \Huniv\ with \solcz\ this gives the relevant contribution to 
the expansion of $\F$ in \expF\ for this case,
\eqn\Fsa{
\F_{g0} (v)= {1\over g!} 
\sum_{j=2}^{g+2} {1\over (j-2)!} \, f_{j-1\, g+3\, j \, g+2}(v) \, .
}
Assuming this form for $\F$ in general then for $p\ge a+3$ \Huniv\ requires, 
for the contributions which involve powers $u^n$ with $n<p$ arising only from the
$T^{(p)}_{pjj,a0}, \, T^{(p)}_{i\, i+k-p-2\, k,a0}$ and $T^{(p)}_{iip,a0}$ terms
in \Hexpall, keeping just the first term in \T, 
\eqn\recura{\eqalign{
& \sum_{j=a+2}^p c^{(p)}_{pjj,a0} \sum_{m=0}^{p-j} {(-1)^m\over m!}(p-j-m)! \, 
u^{j+m-1} f_{j-1+m\, j+1+m\, j+m \, j+m}(v)\cr
&{}+ \sum_{i=p-a+1}^p \sum_{k=a+3}^p c^{(p)}_{i\, i+k-p-2 \, k,a0} 
\sum_{m=0}^{p-k+1} {(-1)^m\over m!}(p-k+1-m)! \cr
\noalign{\vskip - 6pt}
& \hskip 4.8cm {} \times  u^{k+m-2} 
f_{i+k-p-2+m\, k+m \,i+k-p-2+m\, k+m}(v)  \cr
&{}+ \sum_{i=p-a}^{p-1}\,
c^{(p)}_{iip,a0} \,  u^{p-1} f_{i-1\, p+1 \, i \, p }(v) \cr
&{} =  u^{a+1} {1\over a!}
\sum_{j=2}^{a+2} {1\over (j-2)!} \, f_{j-1\, a+3\, j \, a+2}(v) \, . \cr}
}
To analyse \recura\ we consider first all terms proportional to $u^{a+1}$ 
when we obtain
\eqn\abase{\eqalign{
& (p-a-2)! \, c^{(p)}_{p\,a+2\,a+2,a0} \, f_{a+1\, a+3\, a+2 \, a+2}(v) \cr
&{}+ (p-a-2)! \! \sum_{i=p-a+1}^p  c^{(p)}_{i\,i+a-p+1\,a+3,a0} \, 
f_{i+a-p+1\, a+3\, i+a-p+1 \, a+3}(v) \cr
&{}= {1\over a!}
\sum_{j=2}^{a+2} {1\over (j-2)!} \, f_{j-1\, a+3\, j \, a+2}(v) \, . \cr}
}
Applying \hyper\ repeatedly for $f_{j-1\, a+3\, j \, a+2}(v)$ we then get
\eqn\solca{\eqalign{
c^{(p)}_{i\,i+a-p+1\,a+3,a0} = {}& {1\over (p-a-2)! \, a! \, (i+a-p-1)!} \, ,
\quad  i = p-a+1, \dots , p \, , \cr
c^{(p)}_{p\,a+2\,a+2,a0} = {}& {a+1\over (p-a-2)! \, a!^2} \, . \cr}
}
{}From contributions in \recura\ proportional to $u^{k-1} f_{k-1\, k+1\, kk}(v)$,
for $k=a+3,\dots, p-1$, and $u^{k-2} f_{i+k-p-2\, k\, i+k-p-2 \, k}(v)$, 
for $k=a+4, \dots , p, \, i = p-a+1, \dots , p$, we get
\eqn\relcc{
\sum_{m=0}^{k-a-2} {(-1)^m\over m!} \, c^{(p)}_{p\, k-m\,k-m,a0} = 0 \, , \quad
\sum_{m=0}^{k-a-3} {(-1)^m\over m!} \, c^{(p)}_{i\, i+k-p-2-m\,k-m,a0} = 0 \, ,
}
which in conjunction with \solca\ may be solved giving
\eqn\solcaf{\eqalign{
c^{(p)}_{pjj,a0} = {}& {a+1\over (p-a-2)! \, a!^2\, (j-a-2)!} \, , \quad
j = a+2, \dots , p-1 \, , \cr
c^{(p)}_{i\,i+k-p-2\,k,a0} = {}& 
{1\over (p-a-2)! \, a! \, (i+a-p-1)! \,(k-a-3)!}\,,\cr
\noalign{\vskip - 3pt}
&\qquad k=a+3, \dots , p, \, i = p-a+1, \dots , p \, . \cr}
}
Using \solcaf\ the remaining part of \recura\ becomes
\eqn\recurf{\eqalign{
& \bigg ( c^{(p)}_{ppp,a0}  - {a+1 \over (p-a-2)!^2 \, a!^2} \bigg )
\,  f_{p-1\, p+1\, pp }(v)\cr
&{}- {1\over (p-a-2)!^2 \, a!}  
\sum_{j=p-a+1}^p  {1\over (j+a-p-1 )!} \, f_{j-1\, p+1 \, j-1 \, p+1 }(v)  \cr
&{}+ \sum_{i=p-a}^{p-1}\,
c^{(p)}_{iip,a0} \, f_{i-1\, p+1 \, i \, p }(v)  = 0 \, . \cr}
}
With the aid of \hyper\ again we finally obtain for this case
\eqn\solcat{\eqalign{
c^{(p)}_{ppp,a0} = {}& {1\over (p-a-2)!^2 \, a!^2} \, , \cr
c^{(p)}_{iip,a0} = {}& {p-a-1\over (p-a-2)!^2 \, a! \, (i+a-p)!}\,, \quad
i = p-a, \dots , p-1 \, . \cr}
}
For $p=2a+2$ the symmetry conditions \symT\ ensure that in the expansion
\Hexpall\ we may require $c^{(p)}_{pjj,a0}= c^{(p)}_{jjp,a0}$
and $c^{(p)}_{i\, i+k-p-2\, k,a0}$ is symmetric in $i,k$. 
With these constraints instead of \recura\ we have now
\eqn\recurs{\eqalign{
& \sum_{j=a+2}^p c^{(p)}_{pjj,a0} \bigg ( \sum_{m=0}^{p-j} 
{(-1)^m\over m!}(p-j-m)! \,  u^{j+m-1} f_{j-1+m\, j+1+m\, j+m \, j+m}(v) \cr
\noalign{\vskip - 10pt}
& \hskip 8cm {} + u^{p-1} f_{j-1\, p+1 \, j \, p }(v) \bigg ) \cr
&{}+ \!\! \sum_{i,k=a+3}^p \!\! c^{(p)}_{i\, i+k-p-2 \, k ,a0} \sum_{m=0}^{p-k+1} 
{(-1)^m\over m!}(p-k+1-m)! \cr
\noalign{\vskip - 6pt}
& \hskip 4.5cm {} \times  u^{k+m-2}
f_{i+k-p-2+m\, k+m \, i+k-p-2+m\, k+m}(v)  \cr
&{} =  u^{a+1} {1\over a!}
\sum_{j=2}^{a+2} {1\over (j-2)!} \, f_{j-1\, a+3\, j \, a+2}(v) \, . \cr}
}
The solution of \recurs\ is essentially as before giving in this case
\eqn\Sols{\eqalign{
c^{(p)}_{ppp,a0} = {}& {1\over a!^4} \, , \qquad
c^{(p)}_{pjj,a0} = {a+1\over a!^3 \, (j-a-2)!}\,, \quad
j = a+2, \dots , p-1 \, . \cr
c^{(p)}_{i\,i+k-p-2\,k,a0} = {}&
{1 \over a!^2 \, (i-a-3)! \,(k-a-3)!}  \, , \quad i,k = a+3, \dots , p \, . \cr}
}
These results are just as expected from \solcaf\ and \solcat\ after 
substituting $p=2a+2$. The results manifestly satisfy the necessary symmetry
conditions.

For completeness it is also necessary to analyse the other contributions
which are present in crossing symmetric expressions exhibited in \T\ for 
$T^{(p)}_{pjj,a0}, \, T^{(p)}_{i\, i+k-p-2\, k,a0}$ and $T^{(p)}_{iip,a0}$.
The six terms in \T\ form three pairs related by $\sigma\leftrightarrow \tau$
under which our equations are invariant. From those terms proportional to 
$\sigma^{p-2-a}$ we obtain
\eqnn\recurp
$$\eqalignno{
&  \sum_{j=a+2}^p c^{(p)}_{pjj,a0} \, u^{p-1} 
f_{j-1\, p+1\, j \, p}(v)\cr 
&{}+ \! \sum_{i=p-a+1}^p \sum_{k=a+3}^p 
c^{(p)}_{i\,i+k-p-2\, k,a0} \sum_{m=0}^{p-i+1} {(-1)^m\over m!}(p-i+1-m)! \cr
\noalign{\vskip - 4pt}
& \hskip 4.7cm {} \times  u^{i+m-2} 
f_{i+k-p-2+m\, i+m \, i+k-p-2+m\, i+m}(v)  \cr
&{}+ \sum_{i=p-a}^{p-1} c^{(p)}_{iip,a0}\, \sum_{m=0}^{p-i} 
{(-1)^m\over m!}\,(p+1-i-m)!
\,  u^{i+m-1} f_{i-1+m\, i+1+m\, i+m \, i+m}(v)  \cr
&{} = u^{p-1-a} {1\over (p-2-a)!}
\sum_{j=2}^{p-a} {1\over (j-2)!} \, f_{j-1\, p-a+1\, j \, p-a}(v) \cr
&{} = u^{p-1-a} \F_{p-2-a\, 0}(v) \, ,& \recurp\cr}
$$
using \Fsa. The identity shown in \recurp\ is obtained by following the 
identical procedure as in calculations described  above after using
\eqn\symc{
c^{(p)}_{i\, i+k-p-2\, k,a0} \leftrightarrow c^{(p)}_{k\, i+k-p-2\, i,a0} \, , 
\quad c^{(p)}_{pjj,a0} \leftrightarrow c^{(p)}_{jjp,a0} \, , 
\ \ \hbox{for} \ \ a\to p-a-2 \, ,
}
which are easily seen to be a property of the solutions \solcaf\ and \solcat. 
The result \recurp\ is then in accord with expectation from \Huniv.

For the remaining terms we consider those proportional to $\sigma^{p-2-a}
\tau^a$ for which, up to terms of ${\rm O}(u^p)$, we have just
\eqn\sta{
\sigma^{p-2-a}\tau^a u^{p-1} \bigg ( \sum_{j=a+2}^p c^{(p)}_{pjj,a0} \,
f_{j-1\, p+1\, jp}(v) + \sum_{i=p-a}^{p-1} c^{(p)}_{iip,a0} \,
f_{i-1\, p+1\, p\, i }(v) \bigg ) \, .
}
These are identified as required by \Huniv\ with the following term
in the expansion \expF, after using the expressions \solcaf\ and \solcat,
\eqn\Fnew{\eqalign{
\F_{gh}(v)  = 
&  \sum_{j=h+2}^{g+h+1} {h+1 \over g!\, h!^2 \, (j-h-2)!} \,
f_{j-1\, g+h+3 \, j \, g+h+2}(v) \cr
&{}+ \sum_{j=g+2}^{g+h+1} {g+1 \over g!^2 h!\, (j-g-2)!} \,
f_{j-1\, g+h+3 \, g+h+2\, j}(v) \cr
& + {1\over g!^2\, h!^2}\, 
f_{g+h+1\, g+h+3 \, g+h+2\, g+h+2}(v)  \, . \cr}
}
This result is obtained from \sta\ for $g>h$, the corresponding result
for $g<h$ is obtained by using the symmetry under $\sigma \leftrightarrow
\tau$, for $g=h$ it is necessary to use \symc. For $h=0$ \Fnew\ coincides
with \Fsa. Although \Fnew\ is not immediately of the form expected from
\expF\ and \listf, it can be reduced to it by application of \hyper.

With the determination of $\F$ in general in \Fnew\ we can now determine
the remaining coefficients in \Hexpall. For terms proportional to 
$\sigma^a\tau^b$ in \Huniv\ we have, corresponding to \recura, 
\eqn\recurab{\eqalign{
& \sum_{j=a+b+2}^p c^{(p)}_{pjj,ab} \sum_{m=0}^{p-j} {(-1)^m\over m!}(p-j-m)! \, 
u^{j+m-1} f_{j-1+m\, j+1+m\, j+m \, j+m}(v)\cr
&{}+ \sum_{i=p-a+1}^p \sum_{k=a+b+3}^p c^{(p)}_{i\, i+k-p-2 \, k,ab} 
\sum_{m=0}^{p-k+1} {(-1)^m\over m!}(p-k+1-m)! \cr
\noalign{\vskip - 6pt}
& \hskip 4.8cm {} \times  u^{k+m-2} 
f_{i+k-p-2+m\, k+m \,i+k-p-2+m\, k+m}(v)  \cr
&{}+ \sum_{i=p-b+1}^p \sum_{j=a+b+3}^p c^{(p)}_{i\,\, j\, i+j-p-2,ab} 
\sum_{m=0}^{p-j+1} {(-1)^m\over m!}(p-j+1-m)! \cr
\noalign{\vskip - 6pt}
& \hskip 4.8cm {} \times  u^{j+m-2} 
f_{i+j-p-2+m\, j+m \, j+m\, i+j-p-2+m}(v)  \cr
&{}+ \sum_{i=p-a}^{p-1}\,
c^{(p)}_{iip,ab} \,  u^{p-1} f_{i-1\, p+1 \, i \, p }(v) + \sum_{i=p-b}^{p-1}\,
c^{(p)}_{ipi,ab} \,  u^{p-1} f_{i-1\, p+1 \, p \, i }(v) \cr
&{} =  u^{a+b+1}
\bigg ( \sum_{j=b+2}^{a+b+1} {b+1 \over a!\,b!^2 (j-b-2)!} \,
f_{j-1\, a+b+3 \, j \, a+b+2}(v) \cr
& \hskip 1.8cm {} + \sum_{j=a+2}^{a+b+1} {a+1 \over a!^2 b! (j-a-2)!} \,
f_{j-1\, a+b+3 \, a+b+2\,\, j}(v) \cr
&\hskip 1.8cm {} + {1\over a!^2 b!^2}\, 
f_{a+b+1\, a+b+3 \, a+b+2\, a+b+2}(v)  \bigg ) \, . \cr}
}
This may be analysed in a similar fashion to previously. For terms
proportional to $u^{a+b+1}$,
\eqn\abbase{\eqalign{
& (p-a-b-2)! \,\, c^{(p)}_{p\,a+b+2\,a+b+2,ab} \, 
f_{a+b+1\, a+b+3\, a+b+2 \, a+b+2}(v)
\cr &{}+ (p-a-b-2)! \!\! \sum_{i=p-a+1}^p \! c^{(p)}_{i\,i+a+b-p+1\,a+b+3,ab} \, 
f_{i+a+b-p+1\, a+b+3\, i+a+b-p+1 \, a+b+3}(v) 
\cr &{}+ (p-a-b-2)! \!\! \sum_{i=p-b+1}^p \! c^{(p)}_{i\,a+b+3\,i+a+b-p+1,ab}
\,  f_{i+a+b-p+1\, a+b+3\, a+b+3\,i+a+b-p+1}(v) \cr
&{}= \sum_{j=b+2}^{a+b+1} {b+1 \over a!\,b!^2 (j-b-2)!} \,
f_{j-1\, a+b+3 \, j \, a+b+2}(v) \cr
&\quad {} + \sum_{j=a+2}^{a+b+1} {a+1 \over a!^2 b! (j-a-2)!} \,
f_{j-1\, a+b+3 \, a+b+2\,\, j}(v) + {1\over a!^2 b!^2}\, 
f_{a+b+1\, a+b+3 \, a+b+2\, a+b+2}(v) \, . \cr}
}
Using \hyper\ this may be decomposed to give
\eqnn\solcab
$$\eqalignno{
c^{(p)}_{i\,a+b+i-p+1\,a+b+3,ab} = {}& {1\over (p-a-b-2)! \, a! \,b!^2 (i+a-p-1)!}
\, , \quad  i = p-a+1, \dots , p \, ,\cr
c^{(p)}_{i\,a+b+3\,a+b+i-p+1,ab} = {}& {1\over (p-a-b-2)! \, a!^2\,
b!\,(i+b-p-1)!}
\, , \quad  i = p-b+1, \dots , p \, , \cr
c^{(p)}_{p\,a+b+2\,a+b+2,ab} = {}& {a+b+1\over (p-a-b-2)! \, a!^2\,b!^2} \, . &
\solcab \cr}
$$
To obtain \solcab\ we have used the identity $\sum_{m=0}^k (n-1+m)!/m! =
(n+k)!/n\, k!$. For terms proportional to $u^n$ in \recurab\ for
$n=a+b+2, \dots ,p-2$ we get
\eqn\relcab{\eqalign{
\sum_{m=0}^{k-a-b-2} {(-1)^m\over m!} \, c^{(p)}_{p\, k-m\,k-m,ab} = {}& 0 \, ,
\qquad k = a+b+3 , \dots , p-1 \, , \cr
\sum_{m=0}^{k-a-b-3} {(-1)^m\over m!} \, 
c^{(p)}_{i\, i+k-p-2-m\,k-m,ab} ={}&  0 \, , \qquad{}
\cases{i=p-a+1, \dots, p\cr k=a+b+4,\dots ,p} \, , \cr
\sum_{m=0}^{j-a-b-3} {(-1)^m\over m!} \, 
c^{(p)}_{i\, j-m \, i+j-p-2-m,ab} ={}&  0 \, , \qquad
\cases{i=p-b+1, \dots, p\cr j=a+b+4,\dots ,p} \, . 
\cr}
}
Combining with \solcab\ then gives
\eqn\solcabf{\eqalign{
& c^{(p)}_{pjj,ab} = {a+b+1\over (p-a-b-2)! \, a!^2\,b!^2 (j-a-b-2)!} \, , \quad
j = a+b+2, \dots , p-1 \, , \cr
& c^{(p)}_{i\,i+k-p-2\,k,ab} = 
{1\over (p-a-b-2)! \, a! \,b!^2\, (i+a-p-1)! \,(k-a-b-3)!}\,,\cr
\noalign{\vskip - 3pt}
&\hskip 3cm k=a+b+3, \dots , p, \, i = p-a+1, \dots , p \, ,\cr
& c^{(p)}_{i\, j\, i+j-p-2,ab} =  
{1\over (p-a-b-2)! \, a!^2 \,b!\, (i+b-p-1)! \,(j-a-b-3)!}\,,\cr
\noalign{\vskip - 3pt}
&\hskip 3cm j=a+b+3, \dots , p, \, i = p-b+1, \dots , p \, . \cr}
}
For the remaining terms in \recurab\ proportional to $u^{p-1}$ after using the
result \solcabf\ we have
\eqn\recurt{\eqalign{
& \bigg ( c^{(p)}_{ppp,ab}  - {a+b+1 \over (p-a-b-2)!^2 \, a!^2 \,b!^2} \bigg )
\,  f_{p-1\, p+1\, pp }(v)\cr
&{}- {1\over (p-a-b-2)!^2 \, a!\, b!^2}
\sum_{j=p-a+1}^p  {1\over (j+a-p-1 )!} \, f_{j-1\, p+1 \, j-1 \, p+1 }(v)  \cr
&{}- {1\over (p-a-b-2)!^2 \, a!^2\, b!}
\sum_{j=p-b+1}^p  {1\over (j+b-p-1 )!} \, f_{j-1\, p+1 \, p+1 \, j-1}(v)  \cr
&{}+ \sum_{i=p-a}^{p-1}\, c^{(p)}_{iip,ab} \, f_{i-1\, p+1 \, i \, p }(v)
+ \sum_{i=p-b}^{p-1}\, c^{(p)}_{ipi,ab} \, f_{i-1\, p+1 \, p \, i}(v)   
= 0 \, . \cr}
}
This may be solved giving
\eqn\solcabt{\eqalign{
c^{(p)}_{ppp,ab} = {}& {1\over (p-a-b-2)!^2 \, a!^2\, b!^2} \, , \cr
c^{(p)}_{iip,ab} = {}& {p-a-1\over (p-a-b-2)!^2 \, a! \, b!^2\, (i+a-p)!}\,, \quad
i = p-a, \dots , p-1 \, , \cr
c^{(p)}_{ipi,ab} = {}& {p-b-1\over (p-a-b-2)!^2 \, a!^2 \, b!\, (i+b-p)!}\,, \quad
i = p-b, \dots , p-1 \, . \cr}
}
The coefficients satisfy the crucial relations
\eqn\symca{\eqalign{
& c^{(p)}_{\, i\, i+j-p-2\, j,ab} \leftrightarrow c^{(p)}_{j\, i+j-p-2\, i,ab} \, , 
\qquad
c^{(p)}_{\,i\, j\, i+j-p-2,ab} \leftrightarrow c^{(p)}_{\,i+j-p-2\,j\, i,ab} \, , \cr
& c^{(p)}_{iip,ab} \leftrightarrow c^{(p)}_{pii,ab} \, , \qquad 
c^{(p)}_{ipi,ab} \leftrightarrow c^{(p)}_{ipi,ab} \, , \ \ 
\hbox{for} \ \ a\to p-2-a-b \, , \cr}
}
and
\eqn\symcb{\eqalign{
& c^{(p)}_{\,i\, i+j-p-2\, j,ab} \leftrightarrow c^{(p)}_{i+j-p-2\, i\, j,ab} \, ,
\qquad
c^{(p)}_{\,i\, j\, i+j-p-2,ab} \leftrightarrow c^{(p)}_{j\, i \, i+j-p-2,ab} \, , \cr
& c^{(p)}_{iip,ab} \leftrightarrow c^{(p)}_{iip,ab} \, , \qquad
c^{(p)}_{ipi,ab} \leftrightarrow c^{(p)}_{pii,ab} \, , \ \
\hbox{for} \ \ b\to p-2-a-b \, . \cr}
}
There is also a similar relation for $a\leftrightarrow b$ which can be obtained
by combining \symca\ and \symcb.
These relations require for the undetermined coefficient so far
\eqn\solcc{\eqalign{
c^{(p)}_{j+k-p-2\, j \, k ,ab} = {}&
{1\over (p-a-b-2)!^2 \, a! \,b!\, (j+a-p-1)! \,(k+b-p-1)!}\,,\cr
\noalign{\vskip - 3pt}
&\qquad j=p-a+1, \dots , p, \, k = p-b+1, \dots , p \, . \cr}
}
The results \solcabf, \solcabt\ and \solcc\ hence determine the expansion
of $\H^{(p)}$ for general $p$.
The symmetry conditions \symca\ and \symcb\ are necessary to ensure that
the other terms in the expression for $T^{(p)}_{ijk,ab}$, defined in \T,
contribute as required to the terms in $\F$ proportional to $\sigma^g\tau^h$
with $g=p-2-a-b, \, h=b$ and $g=a, \, h=p-2-a-b$. The results given by 
\solcabf, \solcabt\ and \solcc\ also satisfy
\eqnn\symcab
$$\eqalignno{
c^{(p)}_{\,i\, j\, i+j-p-2,aa} = {}& c^{(p)}_{\,i \,i+j-p-2\, j,aa} \, , \qquad
c^{(p)}_{j+k-p-2 \, jk ,aa} = c^{(p)}_{j+k-p-2\,kj ,aa} \, , \cr
c^{(p)}_{\,iip,aa}  = {}& c^{(p)}_{\, ipi,aa} \, , \cr
c^{(p)}_{\,i j\, i+j-p-2,a\, p-2-2a} = {}& c^{(p)}_{i+j-p-2\, j \, i,a\,p-2-2a} \, , 
\ \ c^{(p)}_{\, i \, i+k-p-2 \, k ,a\, p-2-2a} = 
c^{(p)}_{k\, i+k-p-2\,i,a\,p-2-2a} \, , \cr
c^{(p)}_{\,iip,a\, p-2-2a}  = {}& c^{(p)}_{pii,a\, p-2-2a} \, , & \symcab \cr}
$$
which shows that they remain valid in these cases when the symmetry requirements
in \symT\ hold.

\newsec{Cancellation with Free Field Results}

We here endeavour to show that the universal function $\F(u,v;\sigma,\tau)$,
defined by \Huniv\ and given by \expF\ with \Fnew, are precisely what
is required to cancel the universal free field results which have an
expansion of the form \Hcexp. To demonstrate this we need to consider
an expansion in terms of contributions of differing twist,
\eqn\twistF{
(x-\bx)\, u^{g+h+2} \F_{gh}(v) = - (-1)^j \!\!\sum_{j\ge g+h+1}\! \F_{gh,j}(x) \,
g_{1,j+1}(\bx) \, .
}
The leading term is easily seen to be
\eqn\Flead{
\F_{gh,g+h+1}(x) = x^{g+h+3} \F_{gh}(1-x) \, .
}

To obtain an appropriate expression  $\F_{gh,j}(x)$ we follow the basic strategy outlined in
section three by using an integral representation for the hypergeometric functions in
term of  which the basic functions $f_{n_1n_2n_3n_4}$, as defined in \hyper, are
expressed. Considering those which appear in  the solution \Hcexp\  for $\F$ we obtain,
for $n=1,\dots,N$,
\eqn\expfN{\eqalign{
& x^{N+2} f_{n\,N+2\, n+1 \, N+1}(1-x) \cr
\noalign{\vskip -4pt}
&{}=  - (n-1)!N! \bigg ( x ' + (-1)^n
\sum_{s=0}^{N-n} {(N-s)!\, s! \over (N-n-s)!\, (s+n)!} \, x^{s+1} \bigg)
+ \sum_{\ell=0}^{n-1} r^{(n,N)}_\ell \, g_{1,\ell+1}(x) \, , \cr
& (-1)^N x^{N+2}  f_{n\,N+2\, N+1\, n+1}(1-x) \cr
\noalign{\vskip -4pt}
&{}= - (n-1)!N! \bigg ( x  + (-1)^n
\sum_{s=0}^{N-n} {(N-s)!\, s! \over (N-n-s)!\, (s+n)!} \, x'^{s+1} \bigg)
-  \sum_{\ell=0}^{n-1} (-1)^\ell r^{(n,N)}_\ell \, g_{1,\ell+1}(x) \, . \cr}
}
A proof is given in appendix C where an expression for $r^{(n,N)}_\ell$ is obtained.

Using \expfN\ with \Flead\ and \Fnew\ gives
\eqnn\Fgh
$$\eqalignno{
\F_{gh,g+h+1}(x) = {}& (-1)^h \bigg ( {h+1\over h!} \sum_{s=0}^g
{(g+h+1-s)! \over (g-s)!} \, x^{s+1} - {g+1\over g!} \sum_{s=0}^h
{(g+h+1-s)! \over (h-s)!} \, x'^{s+1} \bigg ) \cr
&{} +  \sum_{\ell=0}^{g+h} s_\ell \, g_{1,\ell+1}(x) \, , & \Fgh\cr}
$$
with $s_\ell$ a linear combination of $r^{(n,g+h+3)}_\ell$.
The first term in \Fgh\ is identical with $\K_{gh,g+h+1}(x)$ as given by \cexp\ 
after rewriting in terms of polynomials in $x$ and $x'$. This demonstrates the
cancellation of the leading term of $\F(u,v;\sigma,\tau)$ in a twist
expansion with the corresponding terms found in free field theory. 

The analysis for $\F_{gh,j}(x)$ for $j>g+h+1$ is more involved. Here we
consider just $\F_{00}(v) = f_{1322}(v)$. As shown in appendix C  
the expansion \twistF\ gives for this case
\eqn\Fzero{
\F_{00,j}(x) = {j!^2 \over (2j)!} \, j(j+1) \big ( x + (-1)^j x' \big )
+ {j!^2 \over (2j)!}  \sum_{\ell=0}^{j-1} {\ell!^2 \over (2\ell)!} \, 
c_{j,\ell} \, g_{1,\ell+1}(x) \, ,
}
where
\eqn\cjl{
c_{j,\ell} = (j-\ell)(j+\ell+1) \big ( 1 - (-1)^{j-\ell} \big )  \, .
}
Up to a finite number of partial waves this is identical with $\K_{00,j}(x)$
as given by \Kzero\ which demonstrates the cancellation in this case.
The method used to obtain \Fzero\ is generalisable to other cases but
we have not extended it in general because of algebraic complexity.

\newsec{Conclusion}

We have shown how expressions for four point functions for identical
chiral primary $\half$-BPS operators may be obtained for 
arbitrary $p$, albeit we should note that we require $p\ll N$.
Direct calculations from supergravity assuming the AdS/CFT
correspondence are highly non trivial although our results, if correct, may
suggest that calculations for any $p$ should be possible. The techniques
used here of avoiding $SU(4)$ complications by reducing all expressions
to polynomials may be of assistance. At order $1/N^2$ the supergravity calculations
must reproduce the function $f(x,\alpha)$, for which there are no perturbative
corrections and which is therefore given by free field theory, as well as
the dynamical part contained in $\H(u,v;\sigma,\tau)$. The cancellations between
these contributions suggest that this separation, although following from the 
superconformal Ward identities, is less natural from the viewpoint of supergravity.

It is also possible that the approach
described here might also be applicable for four point functions involving
differing $\half$-BPS operators, although the constraints of crossing symmetry 
which played a crucial role would no longer be present. If feasible this
reflects the remarkable properties of $\N=4$ superconformal theories
in the large $N$ limit.

\bigskip
\noindent
{\bf Acknowledgments}
\medskip
We are grateful to Gleb Arutyunov, Paul Heslop and Emeri Sokatchev for many
helpful conversations.

\vfil\eject
\appendix{A}{Results for Short and Semi-Short Multiplets}

We here illustrate the results of section four by working out more fully 
the contributions of a short and semi-short multiplets. Taking
$\hf (x,\alpha) = \half g_{1,j+k+2}(x) P_i(y)$ then \Gt\ and \QQ\ give
\eqnn\Qshort
$$\eqalignno{
\G^{(j)}_{\Q^0_j,t} = {}& \big ( \de_{t\, j}\, {\ts \sum_n} c_{n,j+k+1} \, 
g_{1,n-1} - c_{t,j+k+1} \, g_{1,j-1} \big ) \, Y_{j-1\, i} \, , \cr
\G^{(j+1)}_{\Q^1_j,t} = {}& \big ( \de_{t\,j+1}\, {\ts \sum_n} c_{n,j+k+1}\, 
g_{1,n-1} - c_{t,j+k+1} \, g_{1,j} \big ) \, \big ( \gamma_{i,1} Y_{j-1\, i+1} 
+ \gamma_{i,-1} Y_{j-1\, i-1} \big ) \, , \cr
\G^{(j+1)}_{\Q^0_{j-1},t} = {}& \big ( \de_{t\,j+1}\, 
{\ts \sum_n} c_{n,j+k+1} \, g_{1,n-1}
- c_{t,j+k+1} \, g_{1,j} \big ) \, Y_{j-2\, i} \, , \cr
\G^{(j+2)}_{\Q^1_{j-1},t} = {}& \big ( \de_{t\,j+2}\,{\ts \sum_n} c_{n,j+k+1}\, 
g_{1,n-1} - c_{t,j+k+1} \, g_{1,j+1} \big ) \, \big ( \gamma_{i,1} Y_{j-2\, i+1}
+ \gamma_{i,-1} Y_{j-2\, i-1} \big ) \, , \cr
\G^{(j+2)}_{\Q^0_{j},t} = {}& \big ( \de_{t\,j+2}\,
{\ts \sum_n} c_{n,j+k+1} \, g_{1,n-1}
- c_{t,j+k+1} \, g_{1,j+1} \big ) \, Y_{j-1\, i} \, , \cr
\G^{(j+3)}_{\Q^0_{j-1},t} = {}& \big ( \de_{t\,j+3}\,
{\ts \sum_n} c_{n,j+k+1} \, g_{1,n-1}
- c_{t,j+k+1} \, g_{1,j+2} \big ) \, Y_{j-2\, i} \, .  & \Qshort \cr}
$$
Hence the definitions  \Gshort\ and \Gsemi\ lead to
\eqn\GC{
\G_{j+r} \big ( \C_{j-1\, i ,k} \big ) = {\ts \sum_{s=0}^3} a_{r,s} \, 
g_{1,j+k+s}\, , \quad \G_{j+k+1+r} \big ( \C_{j-1\, i ,k} \big ) = -  
{\ts \sum_{s=0}^3} a_{s,r} \, g_{1,j-1+s}\, , 
}
for $r=0,1,2,3$ and where
\eqn\ars{\eqalign{
a_{0,s} = {}& c_{j+k+s+1,j+k+1} \, Y_{j-1\, i} \, , \cr
a_{1,s} = {}& c_{j+k+s+1,j+k+1} \big ( \gamma_{i,1} Y_{j-1\, i+1} +
\gamma_{i,-1} Y_{j-1\, i-1} + \gamma_{j,-1} Y_{j-2\, i} \big ) \, , \cr
a_{2,s} = {}& c_{j+k+s+1,j+k+1} \big ( \gamma_{j,-1} (\gamma_{i,1} Y_{j-2\, i+1} +
\gamma_{i,-1} Y_{j- 2\, i-1}) + c_j \, Y_{j-1\, i} \big ) \, , \cr
a_{3,s} = {}& c_{j+k+s+1,j+k+1} \,  \gamma_{j,-1} c_{j+1} \, Y_{j-2\, i} \, . \cr}
}
The contribution of a semi-short multiplet in the conformal partial wave
expansion is then determined by
\eqn\ass{
{\ts\sum_{nm}} a_{nm, j+r\, k+s-r} \big ( \C_{j-1\, i ,k} \big ) Y_{nm} = 
a_{r,s} \, ,
}
which is easily calculated from \ccc\ and \recurY.

{}From \ashort\ and \GC\ for $j>i$ we have setting $k=-1$
\eqn\GB{
\G_{j+r} (\B_{ji} ) = \sum_{s=r+1}^4 b_{r,s} \, g_{1,j+s-1} -
\sum_{s=0}^{r-1} b_{s,r} \, g_{1,j+s-1} \, , \qquad
\gamma_{j,1} b_{r,s} = a_{r,s} - a_{s,r}  \, ,
}
and hence from \Gshort\ and using \Leg, the coefficients are given by
\eqn\gshort{\eqalign{
b_{0,1} = {}& Y_{ji} \, , \cr
b_{0,2} = {}& \gamma_{i,1} 
Y_{j \, i+1} + \gamma_{i,-1} Y_{j \, i-1}+\gamma_{j+1,-1} Y_{j-1 \, i}\, , \cr
b_{0,3} ={}&  c_{j+1} Y_{j\, i} + \gamma_{j+1,-1}\big ( \gamma_{i,1}
Y_{j-1 \, i+1} + \gamma_{i,-1} Y_{j-1 \, i-1}\big )\, , \cr
b_{0,4} ={}& c_{j+2} \gamma_{j+1,-1} Y_{j-1\, i} \, , \cr
b_{1,2} ={}&  \gamma_{j+1,-1} \big ( \gamma_{j,-1} Y_{j-2\, i}
+ \gamma_{i,1} Y_{j-1 \, i+1} + \gamma_{i,-1} Y_{j-1 \, i-1} \big ) \cr
& {} + \gamma_{i+1,1} \gamma_{i,1} Y_{j\, i+2} + 
\gamma_{i-1,-1} \gamma_{i,-1} Y_{j\, i-2} + (c_{i+1} + c_i - c_j) Y_{ji}\, ,\cr
b_{1,3} ={}& \gamma_{j+1,-1} \big ( \gamma_{j,1} ( \gamma_{i,1} Y_{j\, i+1}
+ \gamma_{i,-1} Y_{j \, i-1}) + \gamma_{j,-1} ( \gamma_{i,1} Y_{j-2\, i+1}
+ \gamma_{i,-1} Y_{j-2 \, i-1}) \cr
& \qquad\qquad {} + \gamma_{i+1,1} \gamma_{i,1} Y_{j-1\, i+2} +
\gamma_{i-1,-1} \gamma_{i,-1} Y_{j-1\, i-2} + (c_{i+1} + c_i) Y_{j-1\,i} \big )
\, ,\cr
b_{1,4} ={}& c_{j+2} \gamma_{j+1,-1} \big ( \gamma_{j,-1} Y_{j-2\, i}
+ \gamma_{i,1} Y_{j-1 \, i+1} + \gamma_{i,-1} Y_{j-1 \, i-1} \big ) \, , \cr
b_{2,3} ={}&  \gamma_{j+1,-1}\gamma_{j,-1}  
\big ( \gamma_{j-1,1}\gamma_{j,1} Y_{j\, i} + (c_{i+1} + c_i - c_j) Y_{j-2\, i} \cr
& \qquad\qquad\quad {} + \gamma_{i+1,1} \gamma_{i,1} Y_{j-2\, i+2} +
\gamma_{i-1,-1} \gamma_{i,-1} Y_{j-2\, i-2} \big ) \, ,\cr
b_{2,4} = {}& c_{j+2}\gamma_{j+1,-1}\gamma_{j,-1}  \big ( \gamma_{i,1}
Y_{j-2 \, i+1} + \gamma_{i,-1} Y_{j-2 \, i-1}+\gamma_{j-1,1} Y_{j-1 \, i}\big )
\, , \cr
b_{3,4} = {}& c_{j+1}c_{j+2} \gamma_{j+1,-1}\gamma_{j,-1} Y_{j-2\, i} \, . \cr }
}
{}From the first term in \GB\ and \gshort\ we may then read off the contributions
of the different operators in the $\quar$-BPS multiple $\B_{ji}$ where
$\Delta = 2j+r+s-1$ and $\ell=s-r-1$.

For $i=j$ and $k=-1$, then, as shown in \ashort, \GC\  decomposes into 
contributions from two $\half$-BPS multiplets, $\gamma_{j,1}  ( \G_t (\B_{jj} ) -
\gamma_{j+1,1}\gamma_{j,1} \, \G_t (\B_{j+1\,j+1}) )$. Detailed results can
be easily obtained from \GB\ and \gshort\ using for $n<m$ $Y_{nm}=-Y_{m-1\,n+1}$,
\eqn\GC{
\G_{j+r} (\B_{jj} ) = \sum_{s=r+1}^3 \hb_{r,s} \, g_{1,j+s-1} -
\sum_{s=0}^{r-1} \hb_{s,r} \, g_{1,j+s-1} \, , \qquad r=0,1,2 \, ,
}
where
\eqn\resb{\eqalign{
\hb_{0,1} = {}& Y_{j\,j} \, , \quad \hb_{0,2} = \gamma_{j,-1} Y_{j\,j-1} \, , \quad
\hb_{0,3} = \gamma_{j+1,-1} \gamma_{j,-1} Y_{j-1\,j-1} \, , \cr
\hb_{1,2} = {}& \gamma_{j,-1} \gamma_{j-1,-1} Y_{j\,j-2} \, , \quad
\hb_{1,3} = \gamma_{j+1,-1} \gamma_{j,-1}\gamma_{j-1,-1} Y_{j-1\,j-2} \, , \cr
\hb_{2,3} = {}&\gamma_{j+1,-1} \gamma_{j,-1}{\!\!\!}^2\gamma_{j-1,-1} 
Y_{j-2\,j-2} \, . \cr}
}
Hence the conformal partial wave expansion  for a $\half$-BPS multiplet is 
determined by 
${\sum_{nm}} a_{nm, j+r\, s-r} ( \B_{jj} \big ) Y_{nm} = \hb_{r,s+1}$.
For $j=1$ $\hb_{0,1} = Y_{11}, \, \hb_{0,2} = {1\over 6} Y_{10} , \,
\hb_{0,3} = {1 \over 30}$ and $\hb_{1,2} = \hb_{1,3} = \hb_{2,3} =0$, which
is identical with \Bone.

\appendix{B}{Large $N$ Free Field Results}

We here calculate the leading large $N$ behaviour for the free field contributions for the 
four point function for $[0,p,0]$ chiral primary operators. The generators $\{T_a\}$, $a=1,\dots , N^2$
of $U(N)$ are $N\times N$ matrices  with the following properties
\eqn\UN{
[T_a,T_b]=if_{abc}T_c \, , \quad \tr(T_a T_b) = \half \de_{ab} \,, \quad 
\tr(T_a A) \, \tr(T_a B) = \half \tr(AB) \, , \quad T_a T_a = \half N \, 1 \, .
}
For the purposes of the combinatorics of counting diagrams we define an adjoint 
scalar field with the basic two point function
\eqn\XXC{
\langle X_a X_b \rangle = 2 \de_{ab} \, .
}
Defining the adjoint scalar $X=X_aT_a$ single trace chiral primary operators 
belonging to the $[0,p,0]$ representation correspond to $\tr(X^p)$.

We consider first the two-point function, using the identities \UN,
\eqn\Xtwo{\eqalign{
\big \langle \tr(X^p) \, \tr(X^p)\big  \rangle = {}& 2^p p! \, \tr ( T_{(a_1} \dots T_{a_p)} )
\, \tr ( T_{(a_1} \dots T_{a_p)} ) \cr
\simeq  {}& 2^p p \, \tr ( T_{a_1} \dots T_{a_p} )\, \tr ( T_{a_p} \dots T_{a_1} ) \cr
=  {}& 2^{p-1} p\, \tr ( T_{a_1} \dots T_{a_{p-1}} T_{a_{p-1}} \dots T_{a_1} ) 
= p \, N^p \, ,  \cr}
 }
where in the second line, and also subsequently, $\simeq$ denotes that 
sub-dominant terms for large $N$ have been dropped. In the product of symmetrised 
traces of $T_a$'s, apart from  cyclic permutations of the trace giving the 
factor $p$, only one arrangement of the $T_a$'s, as displayed in the second line, 
gives the leading large $N$ behaviour leading to the factor $N^p$. 
The corresponding planar diagram has index loops $(12)^p$ each generating
a factor $N$.

Corresponding to the three point function of three chiral primary operators 
similarly, taking into account symmetry factors with 
$p_1=i+j, \, p_2=j+k, \, p_3=k+i$,
\eqn\Xthree{\eqalign{
\big \langle  \tr( & X^{p_1}) \, \tr(X^{p_2})\, \tr(X^{p_3})\big  \rangle \cr
= {}& 2^{{1\over 2}(p_1+p_2 + p_3)} p_1! \,p_2!\, p_3! \, {1\over i! j! k!} \cr
\noalign{\vskip -4pt}
&{}\times \tr ( T_{(a_1} \dots T_{a_i}T_{b_1} \dots T_{b_j)} ) 
\, \tr ( T_{(b_1} \dots T_{b_j}T_{c_1} \dots T_{c_k)} ) \,
\tr ( T_{(c_1} \dots T_{c_k}T_{a_1} \dots T_{a_i)} ) \cr
\simeq {}& 2^{{1\over 2}(p_1+p_2 + p_3)} p_1p_2p_3 \cr
\noalign{\vskip -2pt}
&{}\times \tr ( T_{a_1} \dots T_{a_i}T_{b_1} \dots T_{b_j} ) 
\, \tr ( T_{b_j} \dots T_{b_1}T_{c_1} \dots T_{c_k} ) \,
\tr ( T_{c_k} \dots T_{c_1}T_{a_i} \dots T_{a_1} ) \cr
= {}& 2^{{1\over 2}(p_1+p_2 + p_3)-2}\,  p_1p_2p_3 \cr
\noalign{\vskip -2pt}
&{}\times \tr ( T_{a_1} \dots T_{a_i}T_{b_1} \dots T_{b_{j-1}} 
T_{b_{j-1}} \dots T_{b_1}T_{c_1} \dots T_{c_{k-1}} 
T_{c_{k-1}} \dots T_{c_1}T_{a_i} \dots T_{a_1} ) \cr
=  {}& p_1 p_2 p_3 \, N^{{1\over 2}(p_1+p_2 + p_3)-1} \, ,  \cr}
 }
where again we keep only the ordering up to cyclic permutations, which corresponds
to planar diagrams with index loops $(123)(132)(13)^{i-1}(12)^{j-1}(23)^{k-1}$
if $i,j,k>0$. In the extremal case $p_1+p_2=p_3$ or $j=0,\ p_1=i,\ p_2=k$ we have 
instead of \Xthree\ 
$2^{p_3} p_3! \,  \tr ( T_{(a_1} \dots T_{a_i)} )
\, \tr ( T_{(c_1} \dots T_{c_k)} ) \,
\tr ( T_{(c_1} \dots T_{c_k}T_{a_1} \dots T_{a_i)} )$ which for large $N$
reduces to $2^{p_3-1}p_1! p_2 p_3\,
\tr ( T_{(a_1} \dots T_{a_i)} )\, \tr ( T_{c_1} \dots T_{c_{k-1}}
T_{c_{k-1}} \dots T_{c_1}T_{a_i} \dots T_{a_1} ) \simeq  p_1 p_2 p_3 N^{p_3-1}$,
so the final result is unchanged.
The final result in \Xthree\ is of course identical to that of Lee {\it et al} \Sei.

For the four point function of four $\tr(X^p)$ operators we define
\eqn\Xfour{
\big \langle  \tr( X^{p}) \, \tr(X^{p})\, \tr(X^{p})\, \tr(X^{p}) \big  \rangle =
\sum_{i,j,k\ge 0\atop i+j+k=p} I_{ijk} \, ,
}
where $I_{ijk}$ represents diagrams with $i,j,k$ lines linking the external 
vertices, as shown in fig. 1.
{\midinsert
\hfil \epsfxsize=0.6\hsize
\epsfbox{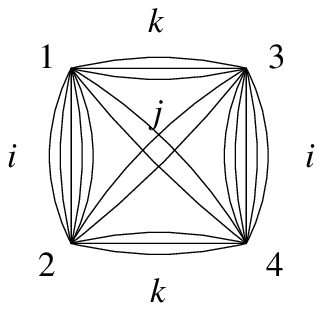} \hfil
\vskip -6pt
{\eightpoint{
\parindent 1cm

{\narrower\smallskip\noindent
Fig. 1 Free field contributions to four point function for $i=5, \, j =2, \, k=3$. 
\smallskip}

\narrower}}
\endinsert}
{}\vskip -12pt
\noindent
With appropriate symmetry factors we have
\eqn\Ifour{\eqalign{
I_{ijk} = 2^{2p}  \! {p!^4\over( i! j! k!)^2} \, &
\tr ( T_{(a_1} \dots T_{a_i}T_{b_1} \dots T_{b_j}T_{c_1} \dots T_{c_k)} ) 
\, \tr ( T_{(c_1} \dots T_{c_k}T_{d_1}\dots T_{d_j}T_{e_1} \dots T_{e_i)} ) \cr
\noalign{\vskip -1pt}
{} \times {}& \tr ( T_{(e_1} \dots T_{e_i}T_{b_1} \dots T_{b_j}T_{f_1} \dots 
T_{f_k)} ) \, \tr ( T_{(f_1} \dots T_{f_k}T_{d_1}\dots T_{d_j}T_{a_1} \dots 
T_{a_i)} ) \, . \cr}
}
If $i\le j \le k$ $I_{ijk}$ and the associated  permutations of $i,j,k$ correspond to the symmetric polynomials $S^{(p)}_{ab}$ in \csb\ where $b=i, \, a=j$. For the 
disconnected diagrams then as in \Xtwo\ we have
\eqn\Idis{
I_{p00}=I_{0p0}=I_{00p} = p^2 N^{2p} \, .
}
If one of $i,j,k$ are zero then we may reduce $I_{ijk}$ to a single trace in the 
same fashion as \Xthree,
\eqn\Izer{\eqalign{
I_{i0k} \simeq {}& 2^{2p-3}p^4\, \tr(  T_{a_1} \dots T_{a_i}T_{c_1} \dots 
T_{c_{k-1}}  T_{c_{k-1}} \dots T_{c_1}T_{e_1} \dots T_{e_{i-1}}  \cr
\noalign{\vskip -1pt}
& \hskip 1.2cm  {} \times   T_{e_{i-1}} \dots T_{e_1}T_{f_1} \dots T_{f_{k-1}} 
T_{f_{k-1}} \dots T_{f_1}T_{a_i} \dots T_{a_1} )  \cr
={}& p^4 N^{2p-2} \, . \cr}
}
In this case the index loops are $(1342)(1243)(13)^{k-1}(24)^{k-1}(12)^{i-1}
(34)^{i-1}$.
For $i,j,k>0$ there are two possibilities
\eqnn\Ione
$$\eqalignno{
I_{ijk} \simeq {}& 2^{2p-3}p^4\big ( \tr( T_{b_1}\dots T_{b_j} T_{a_1} \dots 
T_{a_i} T_{c_1} \dots T_{c_{k-1}}  T_{c_{k-1}} \dots T_{c_1}T_{d_1}\dots T_{d_j}
T_{e_1} \dots T_{e_{i-1}}  \cr
\noalign{\vskip -2pt}
& \hskip 1.3cm  {} \times   T_{e_{i-1}} \dots T_{e_1}T_{f_1} \dots T_{f_{k-1}} 
T_{f_{k-1}} \dots T_{f_1}T_{d_j}\dots T_{d_1}T_{a_i} \dots T_{a_1}
T_{b_j}\dots T_{b_1}  )  \cr
 &\hskip 0.9cm {}+ \tr( T_{d_1}\dots T_{d_j} T_{a_1} \dots T_{a_i}
T_{b_1}\dots T_{b_j}T_{c_1} \dots T_{c_{k-1}}  T_{c_{k-1}} \dots T_{c_1}
T_{e_1} \dots T_{e_{i-1}}  \cr
\noalign{\vskip -2pt}
& \hskip 1.3cm  {} \times   T_{e_{i-1}} \dots T_{e_1}T_{b_j}\dots T_{b_1} T_{f_1} 
\dots T_{f_{k-1}} 
T_{f_{k-1}} \dots T_{f_1}T_{a_i} \dots T_{a_1}T_{d_j}\dots T_{d_1} ) \big )  \cr
={}& 2 p^4 N^{2p-2} \, .    & \Ione \cr}
$$
These results correspond to two independent index loops for the associated 
planar diagrams $(124)(132)(234)(143)(13)^{k-1}(24)^{k-1}
(12)^{i-1}(34)^{i-1}(14)^{j-1}(23)^{j-1}$ and also its reverse 
$(123)(134)(214)(243)(13)^{k-1}(24)^{k-1}(12)^{i-1}(34)^{i-1}(14)^{j-1}(23)^{j-1}$. 
The results \Izer\ and \Ione\ justify \wN, normalising the disconnected 
contribution to 1 by dividing by $p^2 N^{2p}$.

\appendix{C}{Proof of Hypergeometric Identities}

To demonstrate the first result in \expfN\ we use a standard integral
representation for hypergeometric functions,
\eqn\intrep{
x^{N+2} \! f_{n\,N+2\, n+1 \, N+1}(1-x) = (n-1)!\,(N+1)! \int_0^1 
t^n (1-t)^N \, \bigg ( {x \over 1-tx} \bigg )^{N+2} \d t \,.
}
Integrating by parts $N+1$ times gives
\eqnn\intexp
$$\eqalignno{
x^{N+2} f_{n\,N+2\, n+1 \, N+1}(1-x) 
={}& - (n-1)!\, N! \bigg ( x ' + (-1)^n
\sum_{s=0}^{N-n} {(N-s)!\, s! \over (N-n-s)!\, (s+n)!} \, x^{s+1} \bigg) \cr
&{}- (-1)^N (n-1)! \int_0^1 {\d^{N+1} \over \d t^{N+1}} \big (
t^n (1-t)^N \big ) \,  {x \over 1-tx} \, \d t \, . & \intexp \cr}
$$
In the second line we may use the identity \expfa\ to then obtain \expfN\
with
\eqn\rell{
r^{(n,N)}_\ell = (-1)^{\ell+N} {\ell!^2 \over (2\ell)!} \, (n-1)!
\int_0^1 {\d^{N+1} \over \d t^{N+1}} \big (
t^n (1-t)^N \big ) \, P_\ell(2t-1) \, \d t \, ,
}
where it is easy to see that $r^{(n,N)}_\ell=0$ for $\ell\ge n$. The second 
relation in \expfN\ follows directly using the symmetry under $x\to x'$. 
There is no simple evaluation for $r^{(n,N)}_\ell$ but by integration by parts, 
using $P_\ell(2t-1) = (-1)^\ell {1\over \ell!} {\d^\ell \over \d t^\ell}
(t(1-t))^\ell$, 
we may find for $\ell=0,\dots,n-1$, $n=1,\dots N$,
\eqnn\solr
$$\eqalignno{
r^{(n,N)}_\ell = {}& (-1)^\ell {1\over \ell+1}{\ell!\over (2\ell)!} \, 
{(n-1)!^2 n! \, (N+\ell+1)! \over (\ell+n)! \, (n-\ell-1)!}\,
{}_3 F_2 (\ell+1-n,-N,\ell+1;1-n,\ell+2;1 ) \cr
={}& {\ell!^2 \over (2\ell)!}\, (n-1)! \, N! \bigg ( (-1)^\ell - (-1)^n {N \choose n}
\, {}_3 F_2 (-\ell,n-N,\ell+1;-N,n+1;1) \bigg ) \, .  & \solr }
$$
Note that $r^{(1,N)}_\ell = (N+1)! \, \de_{\ell 0}$.
The second formula is sufficient to verify, using \pell,
that the leading terms in an expansion in powers of $x$ on the
right hand side of \expfN\ are cancelled. 

To discuss the expansion of $\F_{00}(v)$ as in \twistF\ we use the same
integral representation as in \intrep\ to first write
\eqn\ftt{
f_{1322}(v) = 2 \int_0^1 { t(1-t) \over (1-tx - t (1-x)\bx )^3} \, \d t \!
= {2 \over 1-x}  \int_0^1 {z(1-z) \over 1 - (1-z)x} \, {1\over (1- z\bx)^3}
\, \d z \, .
}
For application in \twistF\ we use
\eqn\diffz{
2u^2 \, {x- \bx \over (1- z\bx)^3} = - x^2 \bigg ( (1-xz) {\d^2 \over \d z^2 }
- 2x \, {\d \over \d z} \bigg ) {\bx \over 1 - z \bx } \, .
}
The expansion \expfa\ and \twistF\ then gives
\eqn\Fzj{\eqalign{
\F_{00,j}(x) = {}& {j!^2 \over (2j)!} \ xx' \int_0^1 {z(1-z) \over 1 - (1-z)x} \,
\bigg ( (1-xz) {\d^2 \over \d z^2 } - 2x \, {\d \over \d z} \bigg ) 
P_j(2z-1) \, \d z \cr
= {}& - {j!^2 \over (2j)!} \int_0^1 {x^2 \over (1 - (1-z)x)^2} \,
(2z-1) \, {\d \over \d z}  P_j(2z-1) \ \d z \cr
=  {}& {j!^2 \over (2j)!} \, j(j+1) \big ( x + (-1)^j x' \big ) \cr
{}& - {j!^2 \over (2j)!} \int_0^1 {x \over 1 - (1-z)x} \, {\d \over \d z} 
(2z-1) \, {\d \over \d z}  P_j(2z-1) \ \d z  \, . \cr }
}
Inserting the expansion \expfa\ once more gives \Fzero\ with
\eqn\cint{
c_{j,\ell} = \int_0^1 P_\ell(2z-1)  \, {\d \over \d z}
(2z-1) \, {\d \over \d z}  P_j(2z-1) \ \d z \, .
}
It is easy to see that $c_{j,\ell}=0$ for $\ell \ge j$. The integral may be 
evaluated to give \cjl.

More generally we may extend \ftt\ to
\eqn\fttN{
f_{n\,N+2\,n+1\, N+1}(v) = (n-1)!\, (N+1)!  \, {1\over 1-x}  
 \int_0^1 \! {z^n(1-z)  ^N\over (1 - (1-z)x)^n} \, {1\over (1- z\bx)^{N+2}}
\, \d z \, ,
}
and then obtain
\eqn\fjj{\eqalign{
(x-\bx)\,u^{N+1}f_{n\,N+2\,n+1\, N+1}(v)  =  {}& (n-1)! \int_0^1 f^{(n,N)}(x,z) \, 
{\d^N \over \d z^N} {\bx \over 1 - z \bx } \, \d z \cr
= {}& -(-1)^j \sum_j f^{(n,N)}_j (x) \, g_{1,j+1}(\bx) \, , \cr}
}
where
\eqn\fnn{
f^{(n,N)}(x,z) = {x^{N+1} z^{n-1}(1-z)^{N-1} \over (1-(1-z)x)^{n+1}}\,
\big ( n(1-2z) + (n-N) z ( 1 - (1-z) x) \big ) \, .
}
$f^{(n,N)}_j(x)$ may be determined by using \expfa,
\eqn\fnj{
f^{(n,N)}_j(x) = (n-1)! {j!^2 \over (2j)!} \int_0^1 f^{(n,N)}(x,z) \, 
{\d^N \over \d z^N} P_j(2z-1) \, \d z \, ,
}
which is non zero for $j\ge N$.
To evaluate this we express $f^{(n,N)}(x)$ in partial fractions in the form
\eqn\fexp{\eqalign{
& f^{(n,N)}(x,z)  = (-1)^{n} \bigg (\sum_{r=0}^{n} \alpha_r(z) \, 
{1\over r!} {\d^r \over \d z^r}
{x \over 1-(1-z)x} - \sum_{s=0}^{N-n}  \beta_s(z)\, x^{s+1} \bigg ) \, , \cr
& \alpha_r(z) = {N-r-1 \choose n-r} \, z^{n-1}(1-z)^{r-1} \big ( n(1-z) -rz)\big )
\, , \cr
& \beta_s(z) = {N-s-1 \choose n-1} \, z^{n-1}(1-z)^{s-1} \big ( 
s(2z-1) +(N - n) (1-z)\big ) \, , \cr}
}
and then using \expfa\ and integrating by parts $f^{(n,N)}_j(x)$ is
determined as a polynomial in $x$ and $x'$ up to a linear
combination of $g_{1,\ell+1}(x)$ for $\ell=0,\dots, j-N-n+1$ if $j\ge N+n-1$.

As particular cases we have
\eqn\ffj{\eqalign{
f^{(2,2)} (x,z) = {}& z(1-z)(1-2z) \, {\d^2\over \d z^2} {x \over 1-(1-z)x}
\, , \cr
f^{(1,2)}(x,z) ={}& - \Big ( (1-2z){\d \over \d z} + 1 \Big ) 
{x \over 1-(1-z)x} + zx^2 + x \, , \cr}
}
and hence, neglecting terms involving $g_{1,\ell+1}(x)$,
\eqn\fjj{\eqalign{
f^{(2,2)}_j(x) \sim {}& - {j!^2\over (2j)!} \, \half (j-1)j(j+1)(j+2) 
\big ( x + (-1)^jx' \big ) \, , \cr
f^{(1,2)}_j(x) \sim {}& {j^2\over (2j)!} \,\Big (  \half (j-1)j(j+1)(j+2) 
\big ( x - (-1)^j x'\big )\cr
\noalign{\vskip -6pt}
&\hskip 1.2cm  
{} + j(j+1) \big ( x^2 +x +(-1)^j x \big ) - x^2 (1-(-1)^j ) \Big ) \, . \cr}
}
Since $\F_{10}(v) = f_{2433}(v)+ f_{1423}(v)$ from \twistF\ we have 
$\F_{10,j} (x) = f^{(2,2)}_j(x)  + f^{(1,2)}_j(x)$ and the result obtained
from \fjj\ coincides with a direct calculation of $\K_{10,j}(x)$ as in section 5.

\vfil\eject
\listrefs

\bye